\documentclass[12pt,preprint]{aastex}
\usepackage{natbib}

\newcommand{\myemail}{guoyang@nju.edu.cn}

\begin{document}

\title{Twist Accumulation and Topology Structure of a Solar Magnetic Flux Rope}
\author{Y. Guo$^{1,2}$, M. D. Ding$^{1,2}$, X. Cheng$^{1,2}$, J. Zhao$^3$, E. Pariat$^4$}

\affil{$^1$ School of Astronomy and Space Science, Nanjing University, Nanjing 210093, China} \email{\myemail}
\affil{$^2$ Key Laboratory for Modern Astronomy and Astrophysics (Nanjing University), Ministry of Education, Nanjing 210093, China}
\affil{$^3$ Purple Mountain Observatory, Chinese Academy of Sciences, Nanjing 210008, China}
\affil{$^4$ LESIA, Observatoire de Paris, CNRS, UPMC, Universit\'e Paris Diderot, 5 place Jules Janssen, 92190 Meudon, France}

\begin{abstract}
To study the build up of a magnetic flux rope before a major flare and coronal mass ejection (CME), we compute the magnetic helicity injection, twist accumulation, and the topology structure of the three dimensional magnetic field, which is derived by the nonlinear force-free field model. The Extreme-ultraviolet Imaging Telescope on board the \textit{Solar and Heliospheric Observatory} observed a series of confined flares without any CME before a major flare with a CME at 23:02 UT on 2005 January 15 in active region NOAA 10720. We derive the vector velocity at eight time points from 18:27 UT to 22:20 UT with the differential affine velocity estimator for vector magnetic fields, which were observed by the Digital Vector Magnetograph at Big Bear Solar Observatory. The injected magnetic helicity is computed with the vector magnetic and velocity fields. The helicity injection rate was $(-16.47 \pm 3.52) \times 10^{40}~\mathrm{Mx}^2 ~\mathrm{hr}^{-1}$. We find that only about 1.8\% of the injected magnetic helicity became finally the internal helicity of the magnetic flux rope, whose twist increasing rate was  $-0.18 \pm 0.08$ Turns hr$^{-1}$. The quasi-separatrix layers (QSLs) of the three dimensional magnetic field are computed by evaluating the squashing degree, $Q$. We find that the flux rope was wrapped by QSLs with large $Q$ values, where the magnetic reconnection induced by the continuously injected magnetic helicity further produced the confined flares. We suggest that the flux rope was built up and heated by the magnetic reconnection in the QSLs.
\end{abstract}


\section{Introduction} \label{sec:intro}

Magnetic flux ropes play a key role in the models of various solar activities, such as flares, filament/prominence eruptions, and coronal mass ejections (CMEs). They have been thoroughly studied by numerical simulations \citep[e.g.,][]{2000Amari,2004Fan,2007Fan,2004Torok,2005Torok,2010Aulanier} and observations \citep[e.g.,][]{2011Cheng,2013Cheng,2012Zhang,2013Li1,2013Li2}. To drive an eruption, there must be sufficient free energy stored in the magnetic field and forces acted on the erupted object. However, a current-free (potential) magnetic field contains the least amount of magnetic energy in all the possible magnetic configurations given the same magnetic fluxes on the bottom boundary. Therefore, a flux rope with large electric currents is required to exist in the pre-eruptive magnetic field or form during the eruption. The field lines are highly twisted around each other in such a configuration. The loss of equilibrium \citep{1991Forbes} or the torus instability \citep{2006Kliem} ejects the flux rope, whose eruption stretches the overlying magnetic field lines. \citet{2010Demoulin} pointed out that the loss of equilibrium and the torus instability are two different views of the same physical mechanism, which is the Lorentz repulsion force of electric currents with different paths. Or in the magnetohydrodynamic (MHD) point of view, it is the result of the force imbalance between the magnetic pressure in the flux rope and the magnetic tension of the overlying magnetic field.

There are still some key problems awaiting to be answered for the magnetic flux rope eruption. For example, can a magnetic flux rope exist long time before a major flare or CME? How is the non-potential state built up before the flux rope eruption? Does the magnetic reconnection occur in a flux rope?

It matters whether a flux rope exists before an eruption because it determines the initiation mechanism of an eruption. For example, \citet{2005Torok} found that the helical kink instability of a pre-existing twisted flux rope could trigger and initially drive an eruption. Magnetic flux ropes have been found before the onset of flares or CMEs by nonlinear force-free field (NLFFF) extrapolations \citep{2009Canou,2010Canou,2009Savcheva,2010Guo2,2010Guo1,2010Cheng,2011Su} and indicated by extreme-ultraviolet (EUV) and X-ray observations \citep{2009Green,2011Green,2010Liu,2013Patsourakos}. On the other hand, a flux rope can be formed from magnetic arcades during the eruption by magnetic reconnection \citep{1992Moore,2001Moore,1999Antiochos}.

A magnetic flux rope is built up by the line-tied photospheric motions, such as the magnetic flux emergence or the horizontal flows. This process injects the magnetic helicity into the higher solar atmosphere, which increases the twist and kink of a flux rope (self-helicity) and the linkage between different flux ropes (mutual helicity). The magnetic helicity is conserved in an ideal MHD process and changes very slowly in a resistive process. Thus, a flux rope with continuous injection of magnetic helicity inevitably erupts to remove the accumulated helicity. Magnetic helicity injection can be inferred from magnetic field and velocity field observations in the photosphere. Theory and techniques to measure the magnetic helicity and the velocity field have been developed in the past years. \citet{2005Pariat} proposed a new expression $G_\theta$ for the flux density of magnetic helicity. Various optical flow techniques have been proposed and tested with analytical velocity field models and MHD simulations \citep{2002Kusano,2004Welsch,2007Welsch,2004Longcope,2006Georgoulis,2006Schuck,2008Schuck,2008Chae}. However, it is not easy to estimate the reliability of these calculations when they are applied to observations. Therefore, we cross-check the helicity injection in a fast evolving active region with two methods, one from the magnetic field and the velocity field in the photosphere and the other from the NLFFF extrapolation.

With NLFFF extrapolations, one could compute the total relative helicity in a three dimensional volume and the self-helicity of an elementary magnetic flux rope. The comparison between magnetic helicity injection and total relative helicity from force-free fields has been done in a few cases \citep{2007Lim,2010Park,2012Jing}. They all found overall good agreement between the two helicity estimation methods. \citet{2007Lim} assumed linear force-free field models, for which the total relative helicity measurement is mathematically straightforward and well posed. However, there are still problems in dealing with the boundaries in the direct helicity measurement with NLFFF models as did in \citet{2010Park} and \citet{2012Jing}. Four methods to deal with these problems have been proposed by \citet{2011Rudenko}, \citet{2011Thalmann}, \citet{2012Valori}, and \citet{2013Yang1,2013Yang2}. Here, our main purpose is not to quantify the relationship between the magnetic helicity injection from the photosphere and the total relative helicity from NLFFF models. Instead, we compare in detail the former with the self-helicity contained in a magnetic flux rope. Neither the total relative magnetic helicity nor the self-helicity measures the same quantity as the magnetic helicity injection from the photosphere, because the coronal helicity may see variations related to a CME that could not be seen in the helicity flux on the photosphere. However, compared to the total relative helicity, magnetic helicity contained in a magnetic flux rope has a more direct relationship with the trigger mechanism of the eruption.

A traditional view is that magnetic reconnection in a flare occurs in the current sheet tracing behind an erupting flux rope. There is more and more evidence suggesting that it could also occur in the leading edge of a flux rope, both from numerical simulations \citep{2003Amari,2003roussev,2005Torok} and from observations \citep{2003Ji,2009Wang,2011Huang}. \citet{2012Guo} proposed an alternative possibility that the magnetic reconnection could occur inside a flux rope by the internal kink instability. The occurrence of the magnetic reconnection requires the formation of current sheets, which are prone to form at locations where the magnetic linkages change drastically, namely the quasi-separatrix layers (QSLs; \citealt{1995Priest,1996Demoulin,2006Demoulin,2007Demoulin}). In the extreme case, the magnetic linkages are discontinuous, and QSLs degenerate into the separatrix. \citet{1996Demoulin} proposed a norm, $N$, of the Jacobian matrix of the field line mapping to compute the locations of QSLs. However, the norm $N$ is not constant along a magnetic field line. \citet{2002Titov} proposed a new parameter (squashing degree, $Q$) to define QSLs. The squashing degree $Q$ is invariant along a magnetic field line. QSLs are three dimensional volumes where $N$ or $Q$ are large. \citet{2012Pariat} analyzed several methods and proposed the best one to compute two-dimensional (2D) Q maps in the three-dimensional (3D) domain. This magnetic field topology analysis method has been applied to studying coronal sigmoids \citep{2012Savcheva750,2012Savcheva744}.

To solve the above discussed problems, we have to know the 3D kinematic, thermal, and magnetic parameters in the solar atmosphere. The kinematic and thermal parameters can be derived from multi-wavelength imaging and spectral observations. However, the magnetic field information is difficult to observe directly in the higher solar atmosphere above the photosphere, where vector magnetic field has been observed routinely with both ground-based and space-borne instruments. Although big efforts have been made to observe magnetic fields in the chromosphere and corona \citep{1998Judge,2000Lin,2009Kuckein,2012Kuckein}, much information of the 3D magnetic field in the solar atmosphere can only be derived by various magnetic models. Due to the low plasma $\beta$ condition in the solar corona \citep{2001Gary}, the magnetic field, $\mathbf{B}$, is usually modeled by the force-free field model that obeys the equations $\nabla \times \mathbf{B} = \alpha \mathbf{B}$ and $\nabla \cdot \mathbf{B} = 0$. If the torsional parameter $\alpha$ is a constant, the equations describe a linear force-free field. A special case is when $\alpha = 0$, where the model degenerates to a potential field. From observations, it is found that $\alpha$ usually changes in solar active regions \citep{2002Wiegelmann,2002Regnier}. Therefore, the NLFFF model is necessary to model such cases more realistically than the potential or linear force-free field models. The force-free field equations are thus nonlinear and needed to be solved numerically \citep[see the review by][]{2012Wiegelmann}. Several numerical methods to solve the NLFFF equations have been thoroughly tested against analytical, numerical, and observations \citep{2006Schrijver,2008Metcalf,2009DeRosa}.

In this paper, we study the magnetic helicity injection and 3D magnetic field evolutions before an X2.6 class flare that peaked at 23:02 UT on 2005 January 15. An evolving magnetic flux rope extrapolated by the NLFFF model has been reported in \citet{2010Cheng}. Here, we further compute the twist of the flux rope and compare it with the accumulated magnetic helicity. In particular, we study the build up phase of the magnetic flux rope. Since the velocities on the photosphere are much lower than that in the corona, the evolution can be regarded as a quasi-static process. We use a series of NLFFF to approximate this process. In most cases, only one major flux rope exists in a solar active region, and we could use the twist of the flux rope to approximate the self-helicity if it is not highly kinked. Besides, we analyze the magnetic topology, namely the 3D squashing degree (Q) maps, of the evolving magnetic flux rope to study the magnetic reconnection locations. Observations and data analysis are presented in Section~\ref{sec:data}. Results on the comparison between the injected magnetic helicity and the twist evolution and QSLs of the flux rope are described in Section~\ref{sec:resu}. We finally summarize and discuss our findings in Section~\ref{sec:disc}.

\section{Observations and Data Analysis} \label{sec:data}

\subsection{X-ray and EUV Observations}

Two halo CMEs and associated flares were observed in active region NOAA 10720 on 2005 January 15 by the Large Angle and Spectrometric Coronagraph (LASCO; \citealt{1995Brueckner}) and the Extreme-ultraviolet Imaging Telescope (EIT; \citealt{1995Delaboudiniere}), respectively, on board the \textit{Solar and Heliospheric Observatory} (\textit{SOHO}). \citet{2010Cheng} studied the evolution of the flare loops after the peak of the M8.6 flare at 06:38 UT. The second flare (X2.6) peaked at 23:02 UT. The authors found that the flare loops after the first flare were accelerated again and became the envelope field of the second flare and CME, which were driven by the fast evolving flux rope.

\begin{figure}
\includegraphics[width=0.7\textwidth]{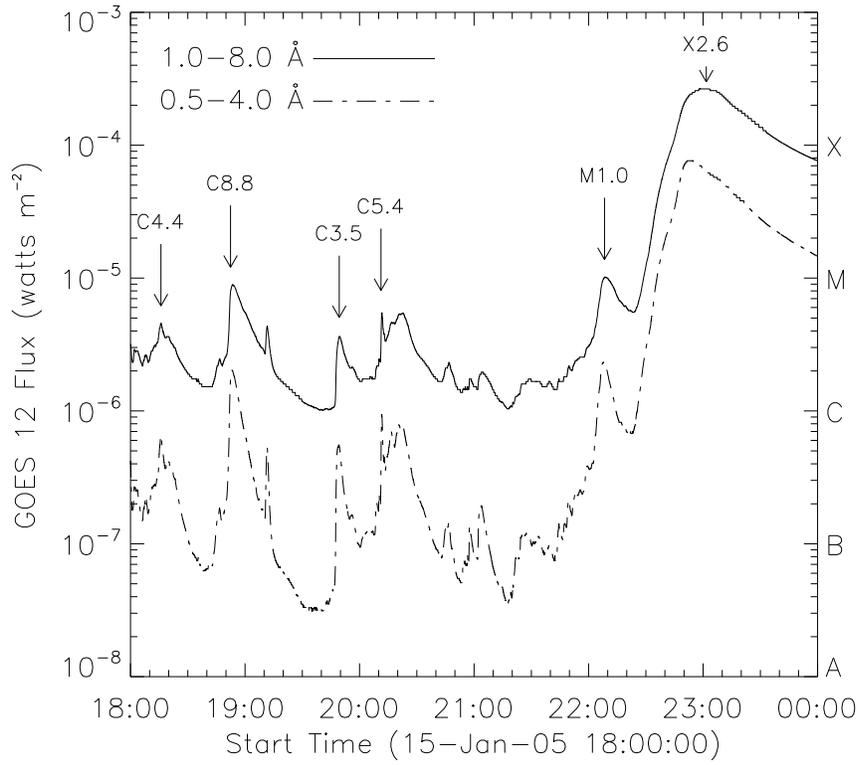}
\caption{GOES soft X-ray fluxes in two wavelength bands from 18:00 to 24:00 UT on 2005 January 15. Solid and dash-dotted lines indicate the fluxes in 1.0--8.0~\AA \ and 0.5--4.0~\AA , respectively.} \label{fig:goes}
\end{figure}

Between the two CMEs and flares, there was no CME detected by \textit{SOHO}/LASCO. However, many flares were observed in active region NOAA 10720 during this period. Here, we only focus on the period between 18:00 to 24:00 UT on 2005 January 15 as shown by the \textit{GOES} soft X-ray fluxes in Figure~\ref{fig:goes}. There were at least five flares before the X2.6 flare in this period in active region NOAA 10720, i.e., the C4.4 (18:16 UT), C8.8 (18:53 UT), C3.5 (19:49 UT), C5.4 (20:11 UT), and M1.0 (22:08 UT) class flares. The time in the parenthesis indicates the peak time of that flare.

\begin{figure}
\includegraphics[width=1.0\textwidth]{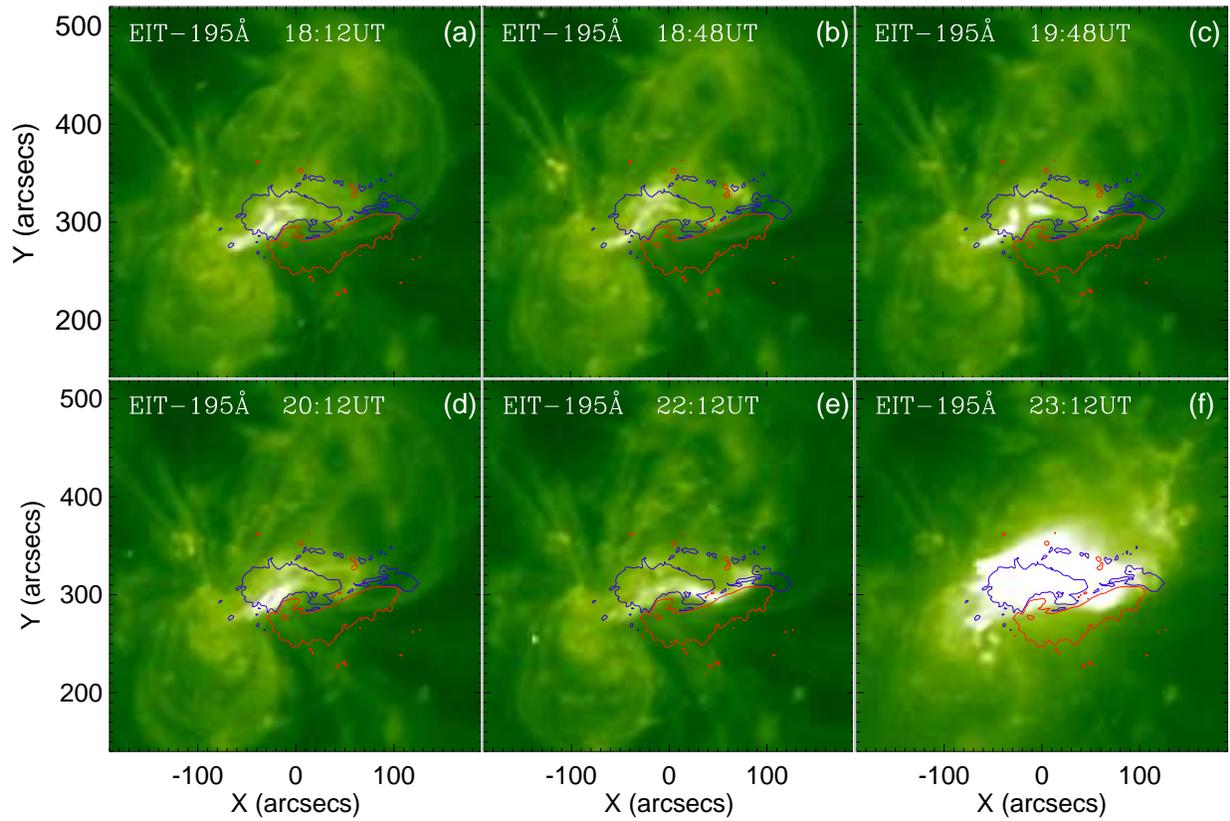}
\caption{\textit{SOHO}/EIT 195 \AA \ images of five confined flares (a)--(e) and one eruptive flare (f) on 2005 January 15. Overlaid red and blue contours denote the positive and negative polarities, respectively, of the line-of-sight magnetic field observed by \textit{SOHO}/MDI.
} \label{fig:195}
\end{figure}

\textit{SOHO}/EIT recorded the aforementioned flares in the 195~\AA \ band. The spatial resolution and temporal cadence of EIT are $2.6''$ per pixel and 12 minutes, respectively. Figure~\ref{fig:195} and an online only movie attached to it displays the evolution of the 195~\AA \ images. We can get two pieces of important information from Figure~\ref{fig:195} and the movie. First, some brightenings appeared intermittently in the core field along the polarity inversion line, as indicated by the five C and M class flares. Secondly, the core field always stayed in the low corona and did not erupt into a higher place. This is because no evidence for a flux rope eruption was found until the X2.6 flare, when the magnetic flux rope eruption led to the second CME.

\subsection{Magnetic Helicity Injection} \label{sec:dhdt}

In the case of the ideal condition that the conductivity approaches infinity, the time ($t$) variation of the magnetic helicity ($H$) can be written as \citep{1984Berger}
\begin{equation}
\frac{\mathrm{d} H}{\mathrm{d} t} = -2 \int\limits_{S} (\mathbf{A}_\mathrm{p} \cdot \mathbf{u}) B_n \mathrm{d} S, \label{eqn:dhdt1}
\end{equation}
where $S$ denotes the boundary surface, $\mathbf{A}_\mathrm{p}$ is the vector potential of the potential field, $\mathbf{u}$ denotes the velocity of the footpoints of flux tubes (namely the flux transport velocity), and $B_n$ is the normal component of the magnetic field. In practical applications to solar events, the magnetic helicity flux through the photosphere dominates those from other surfaces. Therefore, we can integrate the magnetic helicity flux only in the photosphere. The flux transport velocity is defined as \citep{2003Demoulin}
\begin{equation}
\mathbf{u} = \mathbf{v}_t - \frac{v_n}{B_n}\mathbf{B}_t, \label{eqn:u}
\end{equation}
where $\mathbf{v}_t$ and $v_n$ are the transverse and normal components of the velocity, respectively, and $\mathbf{B}_t$ is the transverse component of the magnetic field. The integrand in Equation~(\ref{eqn:dhdt1}) can be recognized as the helicity flux density,
\begin{equation}
G_A(\mathbf{x}) = -2(\mathbf{A}_\mathrm{p} \cdot \mathbf{u}) B_n, \label{eqn:ga}
\end{equation}
where $\mathbf{x}$ is the postion vector. $G_A$ has been adopted to compute the magnetic helicity flux distribution in active regions by many authors \citep{2001Chae,2001Chae2,2004Chae,2002Kusano,2004Kusano,2002Moon1,2002Moon2,2002Nindos,2003Nindos}.

\citet{2005Pariat} found that $G_A$ creates artificial polarities even with simple flows without magnetic helicity injection into the corona. They proposed a new expression for the time variation of the magnetic helicity:
\begin{equation}
\frac{\mathrm{d} H}{\mathrm{d} t} = -\frac{1}{2\pi} \displaystyle\int\limits_{S} \displaystyle\int\limits_{S'} \frac{\mathrm{d} \theta(\mathbf{r})}{\mathrm{d} t} B_n B'_n \; \mathrm{d} S' \; \mathrm{d} S, \label{eqn:dhdt2}
\end{equation}
where
\begin{equation}
\frac{\mathrm{d} \theta(\mathbf{r})}{\mathrm{d} t} = \frac{1}{r^2} \left(\mathbf{r} \times \frac{\mathrm{d} \mathbf{r}}{\mathrm{d} t} \right)_n = \frac{1}{r^2} [\mathbf{r} \times ( \mathbf{u} - \mathbf{u'})]_n , \label{eqn:dthdt}
\end{equation}
where $\mathbf{r} = \mathbf{x} - \mathbf{x'}$ represents the position vector pointing from $\mathbf{x'}$ to $\mathbf{x}$. Equation~(\ref{eqn:dhdt2}) tells us that the magnetic helicity changes due to the rotation between each pair of two infinitesimal magnetic flux tubes \citep{1984Berger2}. The helicity flux density $G_\theta(\mathbf{x})$ is then defined as
\begin{equation}
G_\theta(\mathbf{x}) = -\frac{B_n}{2\pi} \displaystyle\int\limits_{S'} \frac{\mathrm{d} \theta(\mathbf{r})}{\mathrm{d} t} B'_n \; \mathrm{d} S' . \label{eqn:gth}
\end{equation}
\citet{2005Pariat} found that $G_\theta=0$ for the non-rotating motion of a single footpoint. For two magnetic regions with opposite polarities, both $G_A$ and $G_\theta$ have two artificial polarities, but $G_\theta$ is lower by a factor of 10 than $G_A$.

From Equations~(\ref{eqn:dhdt1}) and (\ref{eqn:dhdt2}), we find that it is necessary to measure the velocity field in the photosphere in order to compute the helicity injection. The velocity field can be inferred from a series of magnetic fields by the so-called optical flow techniques, for instance, the local correlation tracking (LCT) method. Using only the normal component of magnetic fields, LCT has been extensively applied to derive the photospheric velocity, $\mathbf{u}_\mathrm{LCT}$, which is assumed to be the horizontal plasma velocity, $\mathbf{v}_t$ \citep{2001Chae,2001Chae2,2004Chae,2002Moon1,2002Moon2,2002Nindos,2003Nindos}. The normal magnetic fields can be constructed from the line-of-sight magnetic fields by assuming that all magnetic fields are vertical. LCT tries to maximize the correlation coefficient of the intensity ($I$) between two images using a prescribed window, which corresponds to \citep{2005Schuck}
\begin{equation}
I(\mathbf{x},t_2) \equiv I[\mathbf{x}-\mathbf{u}_0(t_2-t_1),t_1] .
\end{equation}
If the image intensity is differentiable, $I(\mathbf{x},t)$ satisfies the advection equation:
\begin{equation}
\frac{\partial I}{\partial t} + \mathbf{u}_0 \cdot \nabla I = 0 .
\end{equation}

The normal component of the magnetic induction equation in the ideal condition is \citep{2002Kusano,2004Kusano,2003Demoulin,2004Welsch,2004Longcope}
\begin{equation}
\frac{\partial B_n}{\partial t} + \nabla_t \cdot (B_n \mathbf{v}_t - v_n \mathbf{B}_t) = 0 , \label{eqn:dbdt1}
\end{equation}
or
\begin{equation}
\frac{\partial B_n}{\partial t} + \nabla_t \cdot (\mathbf{u} B_n) = 0, \label{eqn:dbdt2}
\end{equation}
where we have used the expression for the flux transport velocity defined by Equation~(\ref{eqn:u}). Equations~(\ref{eqn:dbdt1}) or (\ref{eqn:dbdt2}) govern the evolution of the magnetic fields on the photosphere. All the newly developed optical flow techniques have considered the induction equation, such as the inductive local correlation tracking method (ILCT; \citealt{2004Welsch}), the minimum energy fit method (MEF; \citealt{2004Longcope}), and the minimum structure reconstruction method (MSR; \citealt{2006Georgoulis}). \citet{2005Schuck} pointed out that the magnetic induction equation is a continuity equation, and the LCT method is inconsistent with it. Therefore, an appropriate method should start with the continuity Equations~(\ref{eqn:dbdt1}) or (\ref{eqn:dbdt2}).

There are two ambiguities in deriving the photospheric velocities from Equations~(\ref{eqn:dbdt1}) or (\ref{eqn:dbdt2}). First, the flux transport vector, which is defined as the product of the flux transport velocity and the normal magnetic field, can be decomposed to \citep{2004Welsch,2004Longcope}
\begin{equation}
\mathbf{u} B_n = B_n \mathbf{v}_t - v_n \mathbf{B}_t = -(\nabla_t \phi + \nabla_t \psi \times \hat{\mathbf{n}}) ,
\end{equation}
where $\hat{\mathbf{n}}$ is the unit vector of the normal direction. The scalar functions $\phi$ and $\psi$ are the inductive and electrostatic potentials. Only the inductive potential $\phi$ can be determined by Equations~(\ref{eqn:dbdt1}) or (\ref{eqn:dbdt2}), while the electrostatic potential $\psi$ could be arbitrary functions. The velocity field cannot be determined uniquely without additional assumptions. For example, \citet{2005Schuck} assumed an affine velocity profile for the velocity model in a small window, where
the velocity is linearly dependent on the local coordinates around the central position. With the affine velocity model, the continuity equation has an analytical solution, thus the velocities can be determined via a least-square method. This optical flow technique is termed as the differential affine velocity estimator (DAVE; \citealt{2006Schuck}). DAVE not only removes the first ambiguity but also guarantees that the dynamics of magnetic fields is consistent with the continuity equation.

The second ambiguity is that the plasma velocity along a field line, $\mathbf{v}_\parallel$, cannot be constrained only by the evolution of the normal magnetogram. The DAVE method has been extended for vector magnetic fields (DAVE4VM; \citealt{2008Schuck}). The plasma velocity is assumed to be a three dimensional affine velocity profile. Under such an assumption, the first and the second ambiguities are removed at the same time. A vector velocity field including the component along a field line can be determined through a time sequence of vector magnetic fields.

\begin{figure}
\includegraphics[width=1.0\textwidth]{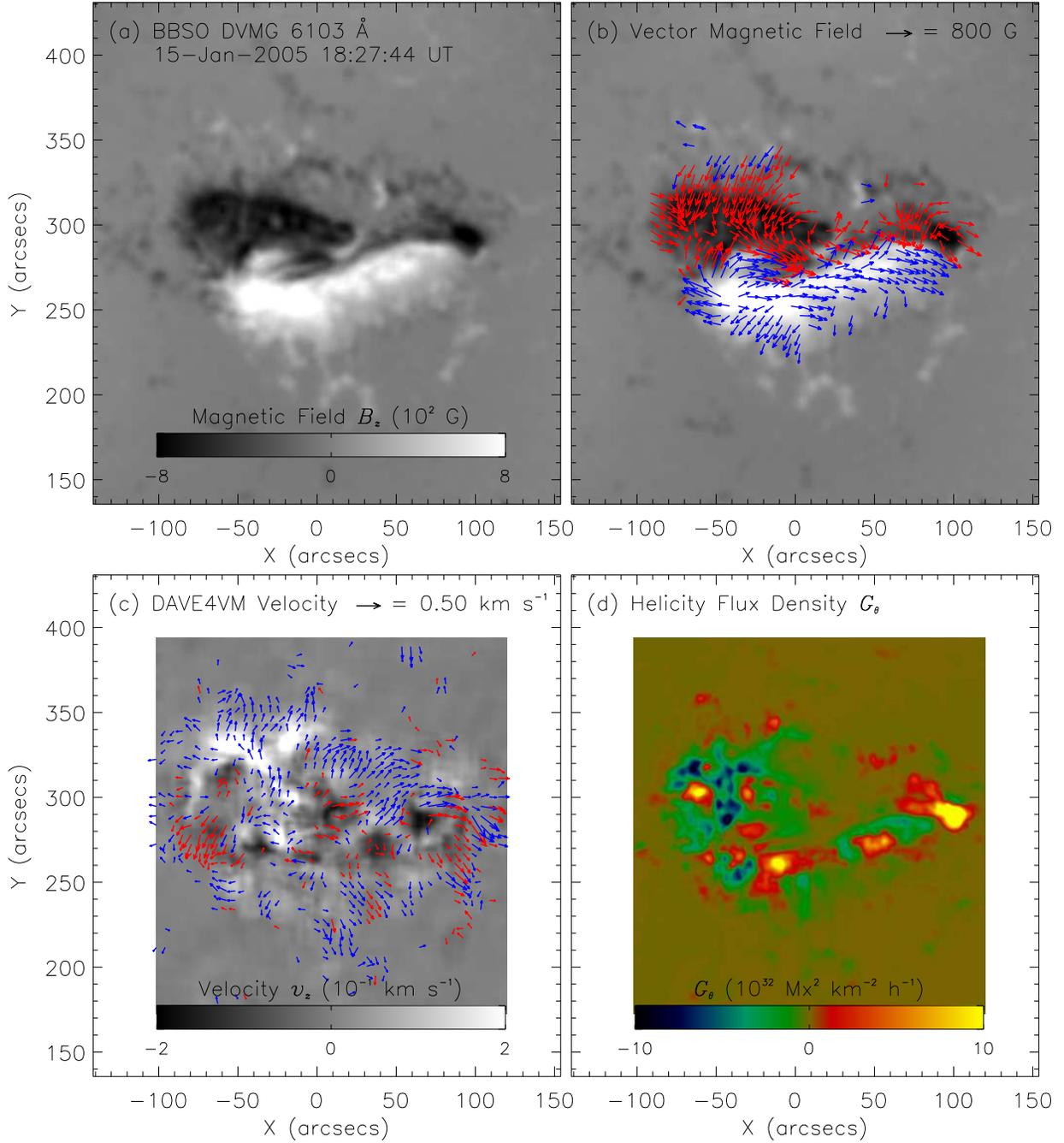}
\caption{(a) Line of sight magnetic field observed by BBSO/DVMG at 18:27 UT on 2005 January 15. (b) Vector magnetic field observed by BBSO/DVMG at 18:27 UT on 2005 January 15. (c) Velocity field derived from DAVE4VM at 18:46 UT on 2005 January 15. (d) Helicity flux density, $G_\theta$, computed by the vector magnetic field and velocity field at 18:46 UT on 2005 January 15.} \label{fig:bvh}
\end{figure}

The Digital Vector Magnetograph (DVMG) at Big Bear Solar Observatory (BBSO) observed a series of vector magnetic fields with a cadence of about 1 minute and a spatial resolution of about $0.6''$. The details of the data process can be found in \citet{2009Jing} and \citet{2010Cheng}. The $180^\circ$ ambiguity of the transverse component of the vector magnetic field is removed by the minimum energy method that minimizes the electric current density and the magnetic field divergence simultaneously \citep{1994Metcalf,2006Metcalf,2009Leka}. Since the active region 10720 was crossing the central meridian and close to the solar disk center, the projection effect is of little influence. The finally processed vector magnetic field at 18:27 UT is shown in Figures~\ref{fig:bvh}a and \ref{fig:bvh}b.

We adopt the DAVE4VM method developed by \citet{2008Schuck} to compute the vector velocity fields through a time sequence of magnetic fields. The window size is selected to be 23 pixels, which is determined by examining the Pearson correlation and slope between $\nabla_t \cdot (\mathbf{u} B_n)$ and $\Delta B_n / \Delta t$ \citep{2008Schuck}. To restrict the amount of computation and resolve small velocities, we select the following vector magnetic fields to analyze in detail at the time of 18:27, 19:04, 19:30, 20:05, 20:38, 21:02, 21:33, and 22:20 UT on 2005 January 15. With each pair of consecutive magnetic fields, we compute the velocity field at the middle time. For example, Figure~\ref{fig:bvh}c shows the velocity derived by DAVE4VM at the middle time (18:46 UT) of 18:27 and 19:04 UT. An averaged vector magnetic field is also constructed by the same pair of magnetic fields. The magnetic helicity flux density, $G_\theta$, is thus obtained by Equation~(\ref{eqn:gth}) and the computed vector magnetic field and velocity field at 18:46 UT. Figure~\ref{fig:bvh}d displays the magnetic helicity flux density at 18:46 UT. The magnetic helicity flux (namely the helicity injection rate or the time variation of the helicity), $\mathrm{d} H / \mathrm{d}t$, is derived by the integration of $G_\theta$ over the photosphere, i.e., by Equation~(\ref{eqn:dhdt2}).

\subsection{Twist of the Flux Rope} \label{sec:twist}

\begin{figure}
\includegraphics[width=0.8\textwidth]{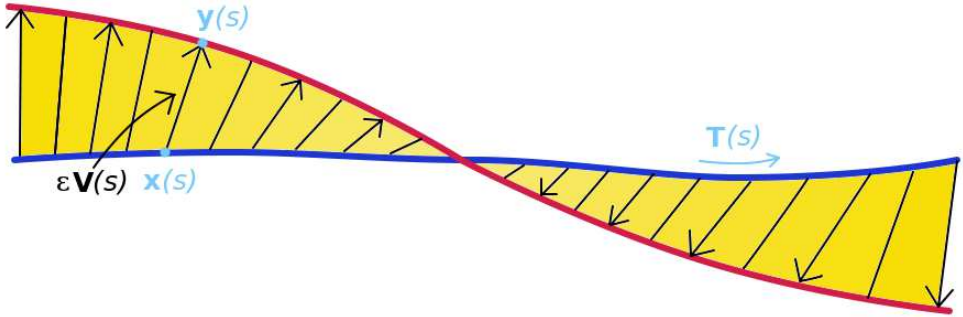}
\caption{An axis curve $\mathbf{x}(s)$ and another arbitrary curve $\mathbf{y}(s)$ showing the way to compute the twist of two curves, where $s$ is the arc length from a reference point on $\mathbf{x}(s)$. $\mathbf{T}(s)$ is a unit vector tangent to $\mathbf{x}(s)$ and $\mathbf{V}(s)$ denotes a unit vector normal to $\mathbf{T}(s)$ and pointing from $\mathbf{x}(s)$ to $\mathbf{y}(s)$, where $\mathbf{y}(s)$ = $\mathbf{x}(s) + \epsilon \mathbf{V}(s)$. This figure is after Figure~1 in \citet{2006Berger}. } \label{fig:twist1}
\end{figure}

\citet{2006Berger} defined the twist between an axis curve and a secondary curve for arbitrary geometries as long as they do not intersect with themselves and are smooth. As shown in Figure~\ref{fig:twist1}, $\mathbf{x}(s)$ denotes a smooth axis curve and $\mathbf{y}(s)$ a secondary curve. $\mathbf{T}(s)$ is a unit vector tangent to $\mathbf{x}(s)$, where $s$ is the arc length from an reference starting point on the axis curve. $\mathbf{V}(s)$ denotes a unit vector normal to $\mathbf{T}(s)$ and pointing from $\mathbf{x}(s)$ to $\mathbf{y}(s)$. Then, the twist density of the secondary curve around the axis is defined by
\begin{equation}
\frac{\mathrm{d} \Phi}{\mathrm{d} s} = \frac{1}{2 \pi} \mathbf{T}(s) \cdot \mathbf{V}(s) \times \frac{\mathrm{d} \mathbf{V}(s)}{\mathrm{d} s} . \label{eqn:twist}
\end{equation}
The total twist is derived by the integration along the axis curve.

Equation~(\ref{eqn:twist}) can be applied to the twisted field lines of the flux ropes, which are constructed by the NLFFF extrapolations with the optimization method \citep{2000Wheatland,2004Wiegelmann}. As an example, Figure~\ref{fig:bvh}b shows the vector magnetic field and its field of view on the photosphere that are used for the NLFFF extrapolation. The flux balance parameter, which is defined as $\int_S B_z \mathrm{d}S/\int_S |B_z| \mathrm{d}S$, is -0.02. It indicates that the bottom boundary is very well flux balanced. Then, we preprocess the vector magnetic field with the method of \citet{2006Wiegelmann} to remove the net magnetic force and torque on the boundary. The preprocessed vector magnetic field is finally used as the bottom boundary for the NLFFF extrapolation.

\begin{figure}
\includegraphics[width=0.8\textwidth]{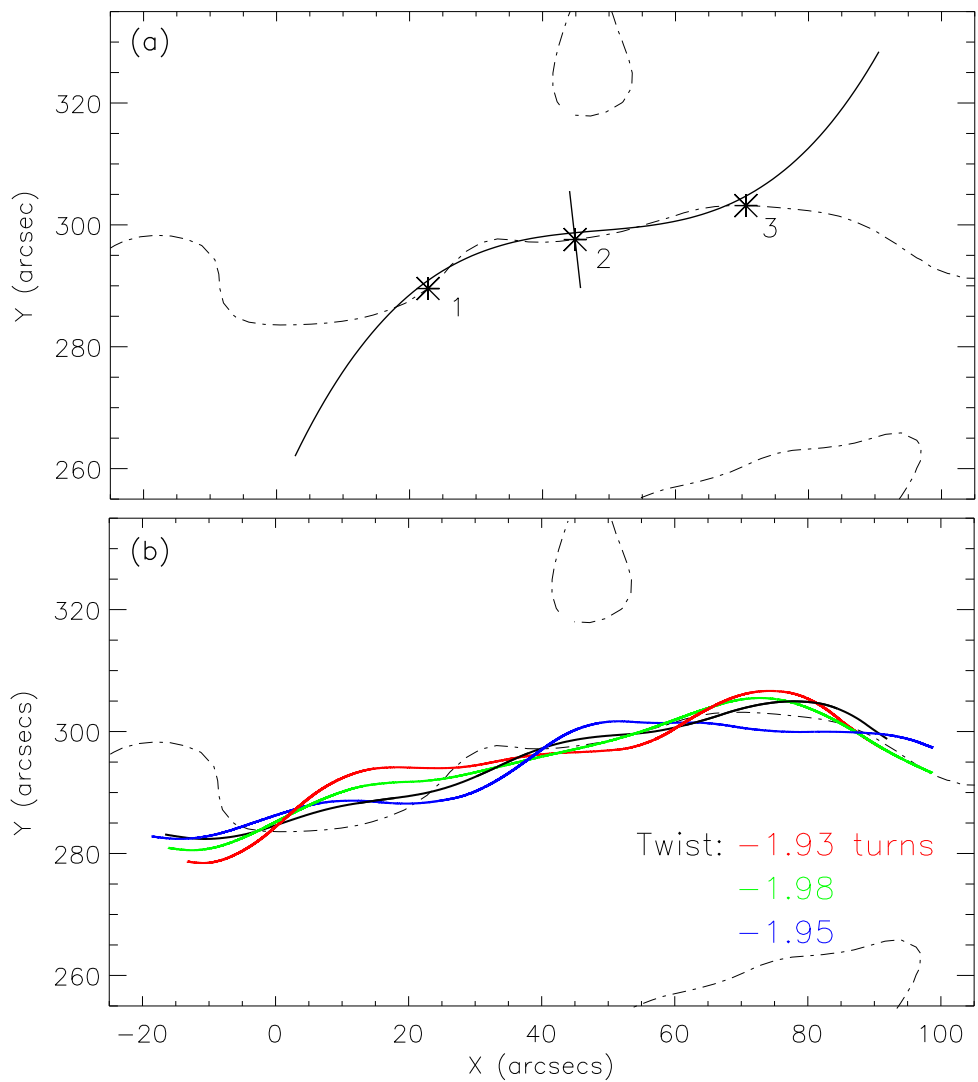}
\caption{(a) Polarity inversion line (dash-dotted curve) associated to the flux rope at 18:27 UT on 2005 January 15. The solid curve indicates a third order polynomial fitted to the polarity inversion line between points 1 and 3. Point 2 is located in the middle of points 1 and 3. The solid line segment at point 2 represents the projection of a $16'' \times 16''$ square, which is perpendicular to the polynomial and to the photosphere. (b) Sample field lines of the flux rope at 18:27 UT on 2005 January 15. The black line is the computed axis of the flux rope. The twist of the sample field lines about the axis is shown in the lower right corner.} \label{fig:twist2}
\end{figure}

To compute the twist of the reconstructed flux rope, we have to determine the axis. From the extrapolated results (as shown in \citet{2010Cheng} and in the following analysis), it is found that the middle section of the flux rope was always horizontal and close to the polarity inversion line. Similar to what has been done in \citet{2010Guo2}, we assume the section of the flux rope between points 1 and 3 in Figure~\ref{fig:twist2}a is horizontal and tangent to the polarity inversion line. An objective function is defined to measure how well the above assumptions are met:
\begin{equation}
F = \int\limits_\mathrm{P1}^\mathrm{P3} |y(x) - y_\mathrm{p}(x)| \mathrm{d} x + \int\limits_\mathrm{P1}^\mathrm{P3} |z(x) - z_\mathrm{a}| \mathrm{d} x, \label{eqn:f}
\end{equation}
where $y(x)$ and $z(x)$ are the Cartesian coordinates of a magnetic field line, $y_\mathrm{p}(x)$ is the $y$-component of the polarity inversion line, and $z_\mathrm{a}$ is the average height of a magnetic field line. The integration is done between points 1 and 3. The axis field line should have a minimum objective function $F$.

To calculate the starting points for integrating the sample magnetic field lines and determining the axis, we fit the polarity inversion line between points 1 and 3 as shown in Figure~\ref{fig:twist2}a with a third order polynomial. Points 1 and 3 are selected in the middle part of the magnetic flux rope with a certain degree of freedom. We will test the errors in computating the twists caused by the positions of points 1 and 3 in the following analysis. A square surface of $16'' \times 16''$ is selected perpendicular to the third order polynomial in the middle of points 1 and 3 (at point 2) and perpendicular to the bottom surface. The area of the square is selected to be large enough to include all the magnetic field lines of the magnetic flux rope and small enough to miminize the computation efforts. The bottom edge of the square is on the photosphere, and the projection of the square is shown as the line segment in Figure~\ref{fig:twist2}a. Then, we integrate some field lines starting from $161 \times 161$ sample points uniformly distributed on the square. The field lines are computed in the computation box until they reach the boundary. Finally, the axis is determined as the magnetic field line with minimum objective function $F$ defined in Equation~(\ref{eqn:f}). The axis is shown as the dark field line in Figure~\ref{fig:twist2}b.

The sample field lines of the magnetic flux rope is determined by two criteria. Namely, they all pass through the $16'' \times 16''$ square and are longer than the axis. The twist density between a sample magnetic field line and the axis is computed with Equation~(\ref{eqn:twist}). The total twist is then derived by the integration of the twist density along the axis. The twists of some sample field lines are noted in Figure~\ref{fig:twist2}b. We quantify the twist of the flux rope as the average twist of the sample field lines of the magnetic flux rope. The average twist is about $\sim 1.90 \pm 0.27$ turns, or $(3.80 \pm 0.54) \pi$ in radian, for the flux rope at 18:27 UT on 2005 January 15. The error is estimated as the standard deviation of the twists of the sample field lines.

The error can also be caused by the determination of the axis, which depends on the locations of points 1 and 3 in Figure~\ref{fig:twist2}a. If we assume the $x$-coordinates of points 1 and 3 are $x_1$ and $x_3$, respectively, the following four cases are considered to compute the twist error caused by the determination of the axis. We chose four pairs of points 1 and 3 on the polarity inversion line and their $x$-coordinates are located at \{$x_1 + 5.0'', x_3 + 5.0''$\}, \{$x_1 - 5.0'', x_3 - 5.0''$\}, \{$x_1 + 5.0'', x_3 - 5.0''$\}, and \{$x_1 - 5.0'', x_3 + 5.0''$\}. The average and standard deviation of the twists of the four cases at 18:27 UT are -1.90 and 0.03 turns, respectively. Compared to $1.90 \pm 0.27$ turns derived from the original case, the average twist of the four test cases are the same, and the standard deviation is much smaller. We have also checked the results for the other 7 samples, whose results are similar to this one at 18:27 UT. Therefore, the method to determine the axis of the flux rope is very robust. The twist does not depends sensitively on the selection of the choice of points 1 and 3.

\section{Results} \label{sec:resu}

\subsection{Accumulated Helicity and Twist Evolution}

\begin{figure}
\includegraphics[width=1.0\textwidth]{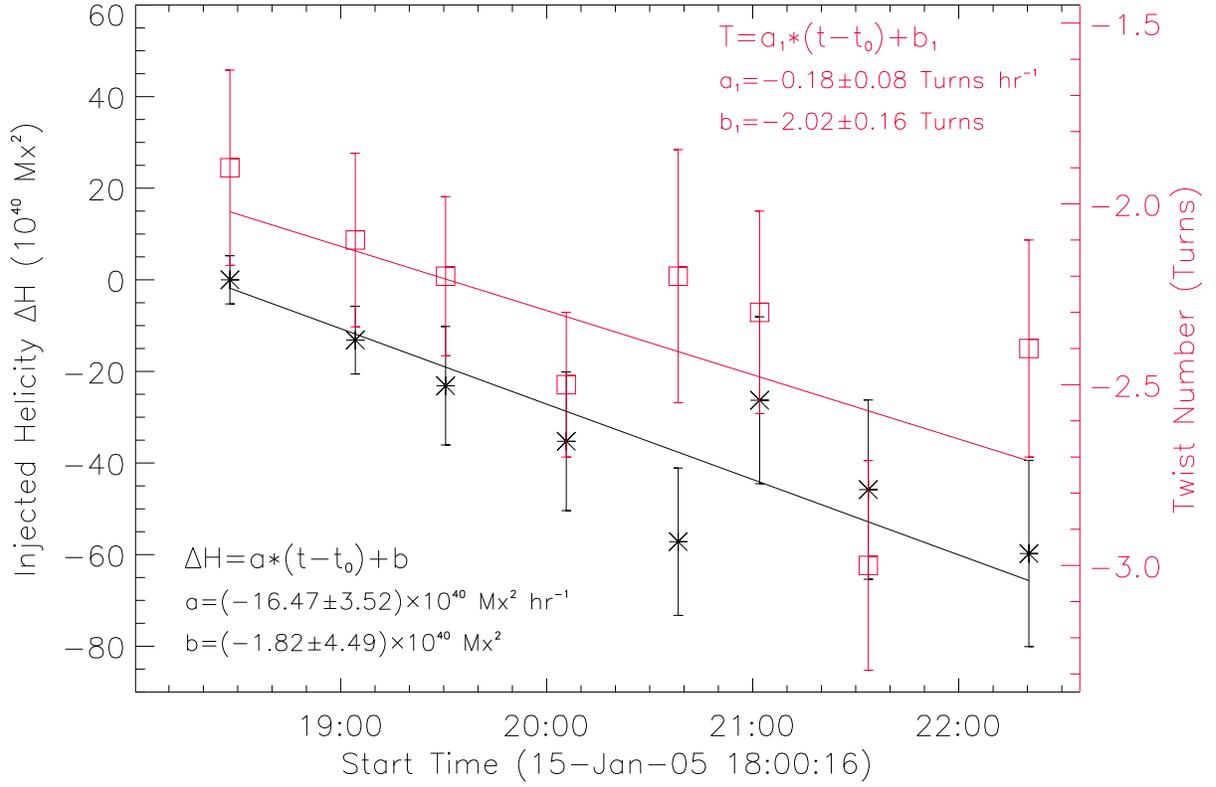}
\caption{Evolutions of the accumulated helicity (asterisk) and the twist (square). Refer to the text for details on the error estimations. The solid lines represent the linear fittings to the accumulated helicity and twist evolutions. The time $t$ is in the unit of hour and referred to $t_0$ at 18:27 UT.} \label{fig:ht}
\end{figure}

Using the methods described in Sections~\ref{sec:dhdt} and \ref{sec:twist}, we compute the helicity injection rate and twist of the flux rope at the time of 18:27, 19:04, 19:30, 20:05, 20:38, 21:02, 21:33, and 22:20 UT on 2005 January 15. The injected magnetic helicity $\Delta H$ is evaluated by
\begin{equation}
\Delta H_{t+\Delta t} = \Delta H_t + \left. \frac{\mathrm{d} H}{\mathrm{d} t} \right |_{t+\frac{1}{2}\Delta t} \Delta t , \label{eqn:delh}
\end{equation}
with $\Delta H_{t_0} = 0$, where $t_0$ is a reference time. Here, we select $t_0$ to be 18:27 UT, i.e., the first moment we extrapolate the 3D magnetic field. The time evolution of $\Delta H$ is plotted in Figure~\ref{fig:ht}, which shows that the negative helicity is injected during the studied period. With a linear fitting, we find that the helicity injection rate is $(-16.47 \pm 3.52) \times 10^{40}~\mathrm{Mx}^2 ~\mathrm{hr}^{-1}$.

The errors of the injected helicity, $\Delta H$, are estimated as follows. Equations~(\ref{eqn:dhdt2}) and (\ref{eqn:dthdt}) indicate that the errors come both from the magnetic field measurements and the velocity field computation. For the velocity field, the spatial alignment between two consecutive magnetic fields also affects the accuracy derived by DAVE4VM, besides the magnetic field measurement errors. For simplicity, we consider the two effects one by one. The first source of the errors of the helicity injection rate ($\mathrm{d}H/\mathrm{d}t$) can be estimated by considering the errors in the magnetic field measurements. Some artificial errors in the normal distribution with the standard deviation of 2 G for $B_z$ and 10 G for $B_x$ and $B_y$ are added to the vector magnetic field, with which the velocity field is derived using the DAVE4VM method. Here, the errors for $B_z$ and for $B_x$ and $B_y$ are determined by the DVMG sensitivity. Then, the helicity injection rate, $\mathrm{d}H/\mathrm{d}t$, is computed with Equations~(\ref{eqn:dhdt2}) and (\ref{eqn:dthdt}). We repeated the above process 10 times. The error for $\mathrm{d}H/\mathrm{d}t$ is estimated as the standard deviation of the 10 results. We computed the errors of $\mathrm{d}H/\mathrm{d}t$ for all the 7 middle time points of the 8 selected samples. The maximum and minimum errors are $0.95 \times 10^{40}$ and $0.23 \times 10^{40} ~\mathrm{Mx}^2 ~\mathrm{hr}^{-1}$, respectively.

The second source of the errors of the helicity injection rate ($\mathrm{d}H/\mathrm{d}t$) is estimated by considering the spatial alignment between two consecutive magnetic fields. We assume the upper limit of the alignment error is one pixel. Then, the second magnetic field is shifted one pixel upward, downward, leftward, and rightward, respectively, each of which can be paired with the first magnetic field and yield a new velocity field. Therefore, we get four more velocity fields in addition to the one computed by the originally aligned magnetic fields. Next, the helicity injection rate, $\mathrm{d}H/\mathrm{d}t$, is computed with Equations~(\ref{eqn:dhdt2}) and (\ref{eqn:dthdt}) based on the five velocity fields. We further estimate the standard deviation of the five results as three times the error of the helicity injection rate. The errors are also computed for the other 6 middle time points. The maximum and minimum errors are $3.53 \times 10^{40}$ and $1.74 \times 10^{40} ~\mathrm{Mx}^2 ~\mathrm{hr}^{-1}$, respectively. Since the errors of $\mathrm{d}H/\mathrm{d}t$ caused by the magnetic field measurement errors are much smaller than that caused by the spatial alignment, we only consider the latter in estimating the errors of the injected helicity, $\Delta H$. The errors are finally computed with the error propagation formula for the addition when Equation~(\ref{eqn:delh}) is used to compute $\Delta H$. If we include the magnetic field measurement errors, it would only increase the errors of $\Delta H$ slightly. However, the following conclusion does not change.

The twists of the flux rope at all the 8 selected time points are computed with the method described in Section~\ref{sec:twist}. The twist error at each time is estimated as the standard deviation of the twists of the sample field lines. Figure~\ref{fig:ht} shows the time evolution of the twist, whose absolute value increases with the time. The negative sign indicates that the twist is left-handed. A linear fitting to the twist time evolution suggests that the twist increasing rate is $-0.18 \pm 0.08$ Turns hr$^{-1}$.

As described in \citet{2000Priest}, the magnetic helicity of a magnetic field can be divided into the twist and kink of elementary flux tubes and the linkage between them, which are deemed as the self-helicity ($H_s$) and mutual helicity ($H_m$), respectively:
\begin{equation}
H = \sum\limits_{i=1}^{N} H_{si} + \sum\limits_{i,j=1(i<j)}^{N} H_{mij},
\end{equation}
where the summation runs over $N$ flux tubes. The self-helicity of an elementary flux tube is
\begin{equation}
H_{si} = T_i F_i^2, \label{eqn:hsi}
\end{equation}
where $T_i$ is the twist in the unit of turns and $F_i$ is the axial magnetic flux of the $i$th magnetic flux tube. In our case, there is only one major twisted flux tube (or flux rope). The ratio between the radius and the length of the flux rope is less than $1/15$ as shown in Figure~\ref{fig:twist2}, which shows that the radius is less than $8''$ and the length is $\sim 120''$. Therefore, the flux rope is relatively thin, and the magnetic helicity of the flux rope can be estimated as the twist since the axis of the flux rope is not highly kinked.

The axial magnetic flux of the flux rope is estimated as follows. We cut a section perpendicular to the flux rope at point 2 as shown in Figure~\ref{fig:twist2} and as described in Section~\ref{sec:twist}. The magnetic flux of the flux rope is defined as the integration of the magnetic flux of the field lines belonging to the flux rope through the above defined cross section. The flux rope field lines have been defined in Section~\ref{sec:twist} as the axis field line and those sample field lines that are longer than it. We compute the axial magnetic field and the magnetic fluxes for all the 8 samples. The mean values of the 8 results are 520 G and $1.28 \times 10^{20}$ Mx, respectively. We have derived that the twist increasing rate is $-0.18 \pm 0.08$ Turns hr$^{-1}$. Following Equation~(\ref{eqn:hsi}), the corresponding self-helicity increasing rate is $(-0.29 \pm 0.13) \times 10^{40} ~\mathrm{Mx}^2 ~\mathrm{hr}^{-1}$, which is about 1.8\% of the total helicity flux, $(-16.47 \pm 3.52) \times 10^{40}~\mathrm{Mx}^2 ~\mathrm{hr}^{-1}$, injected from the bottom boundary.

\subsection{QSLs of the Flux Rope}

\citet{2012Pariat} analyzed three methods to compute the 2D squashing degree (Q) maps in the 3D domain. They proposed that the third method (their Equations (12) to (22)) is the most accurate, which we adopt to do the following analysis. The normal direction of the 2D cut points to the right. It moves from left to right to scan the 3D domain. To save the computation resource and to increase the spatial resolution, we only compute the QSLs in a sub-domain of the NLFFF computation box. For the time at 18:27 UT, the sub-domain is selected as $x \in [-41.5'',108.5'']$, $y \in [269.7'',325.1'']$, and $z \in [0.0'',25.4'']$, where $x$ and $y$ are the heliocentric coordinates and $z$ represents the height referred to the solar surface. This sub-domain is resolved by $325 \times 120 \times 55$ grid points. Therefore, the spatial resolution to compute the QSLs is 5 times the original one to resolve the NLFFF extrapolation.

\begin{figure}
\includegraphics[width=0.9\textwidth]{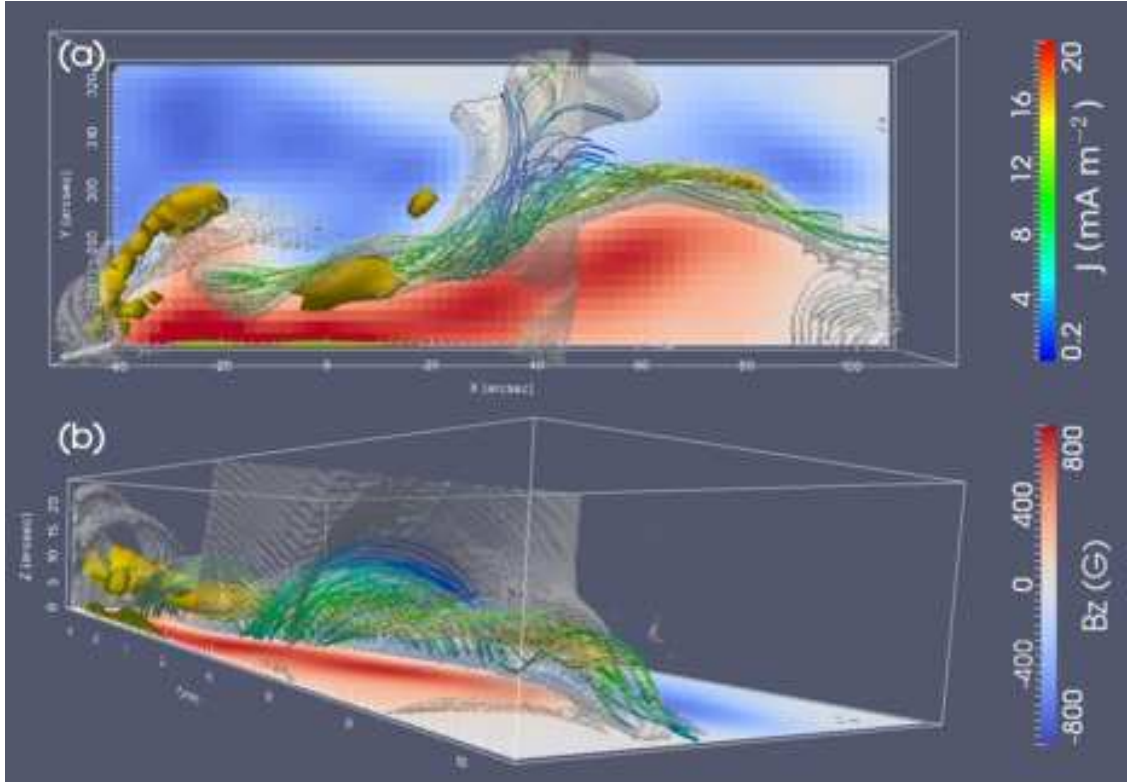}
\caption{Three dimensional contours of the QSLs and electric current density at 18:27 UT on 2005 January 15. Red and blue image represents $B_z$ on the bottom surface with the color scale shown in the lower right color bar. Solid lines depict the magnetic field line, which is rendered according to the magnitude of the electric current density ($J$). Golden surfaces represent the contours of the electric current density at $J = 16~\mathrm{mA~m}^{-2}$. Semitransparent white surfaces represent the contours of the squashing degree at $Q = 10^4$. (a) Top view. (b) Side view.} \label{fig:qsl_j1}
\end{figure}

The 3D distribution of the squashing degree $Q$ is derived by the scanning of the 2D vertical cuts along the $x$-axis. The maximum value is about $10^{13}$. It is visualized by a 3D contour at $Q = 10^4$ as shown in Figure~\ref{fig:qsl_j1}. The most prominent feature of the QSLs is that they go along and wrap the flux rope. Similar to the helical structure of the twisted magnetic field lines, the hollow shell of the QSL contours along the flux rope also have a helical shape. These QSLs are associated with bald patches, where the magnetic field lines touch the bottom surface on the polarity inversion line and concave up. The magnetic field line mappings at the bald patches are discontinuous; therefore, the $Q$ values are extremely large in these regions. But due to the finite spatial resolutions, $Q$ cannot be infinite in practical numerical computations. Also, as a generalization to true separatrix surfaces, such as those at the bald patches, QSLs along the flux rope include more features. For example, the extended QSL contours at about $x \in [40'',60'']$ represent some magnetic field lines bifurcating from the flux rope. The magnetic field line mappings are not necessarily discontinuous, but change drastically.

Besides the QSLs with large $Q$ values along the main flux rope, some are distributed in the eastern part detached from the flux rope at about $x \in [-40'',0'']$ as shown in Figure~\ref{fig:qsl_j1}a. They belong to some sheared and twisted field lines to the east of the main flux rope. In fact, the \textit{SOHO}/EIT 195~\AA \ image indicates the existence of these QSLs as shown in Figures~\ref{fig:195}a and \ref{fig:195}b, which shows that the brightening extended longer and more to the east than the main flux rope in Figure~\ref{fig:qsl_j1}a. However, due to the limitation of the spatial resolution of \textit{SOHO}/EIT, it is not clear if the brightening feature disconnected in the middle part where the magnetic field lines disconnected.

\begin{figure}
\includegraphics[width=0.6\textwidth]{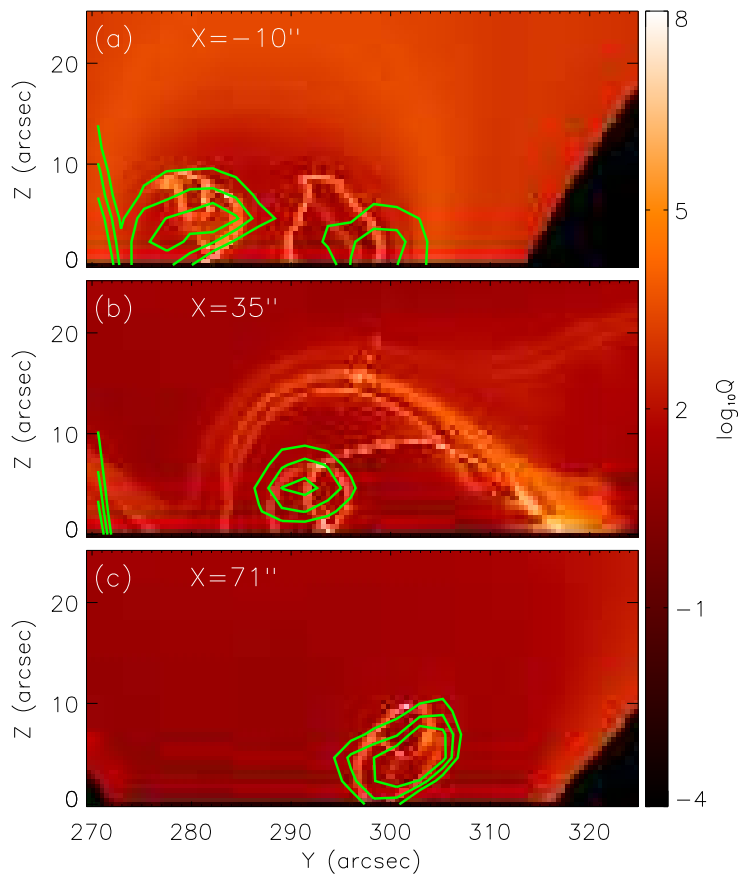}
\caption{Distributions of the logarithm of the squashing degree ($\log_{10}Q$) and the magnitude of the electric current density ($J$) on three selected vertical cuts at 18:27 UT on 2005 January 15. The three cuts are perpendicular to the $x$-axis and at $x=-10''$, $35''$, and $71''$ for panels (a), (b), and (c), respectively. The images display the distributions of the squashing degree on the cuts, and the contours display that of the electric current density. The levels of the contours are 8, 10, and 12 mA~m$^{-2}$ for the solid line from outmost to innermost.} \label{fig:qsl_j2}
\end{figure}

In Figure~\ref{fig:qsl_j1}, we also plot the 3D contours of the electric current density, $J = |\mathbf{J}|$, where $\mathbf{J} = \frac{1}{\mu_0} \nabla \times \mathbf{B}$. Similar to the distribution of the QSLs, the electric currents are also along the magnetic flux rope. It is the basic property of the NLFFF model, which requires that the electric current density, $\mathbf{J}$, is parallel to the magnetic field, $\mathbf{B}$. However, there is one major difference between the distributions of the QSLs and the electric current density. It is a hollow shell for the QSLs, but a solid body for the electric current density. To clearly display this feature, we plot the distributions of the QSLs and the electric current density on three selected cuts, which are both perpendicular to the $x$-axis, as shown in Figure~\ref{fig:qsl_j2}. It is found that the QSLs, where the squashing degree is large, are self-closed or self-intersected at some places. But the electric current density is large at a center part, and the magnitude decreases from the center to the periphery. From Figure~\ref{fig:qsl_j2}, we also find that the places where $J$ is the largest is also where the QSLs appears. But there are also places where QSLs appear while $J$ is not necessarily large.

\begin{figure}
\includegraphics[width=0.65\textwidth]{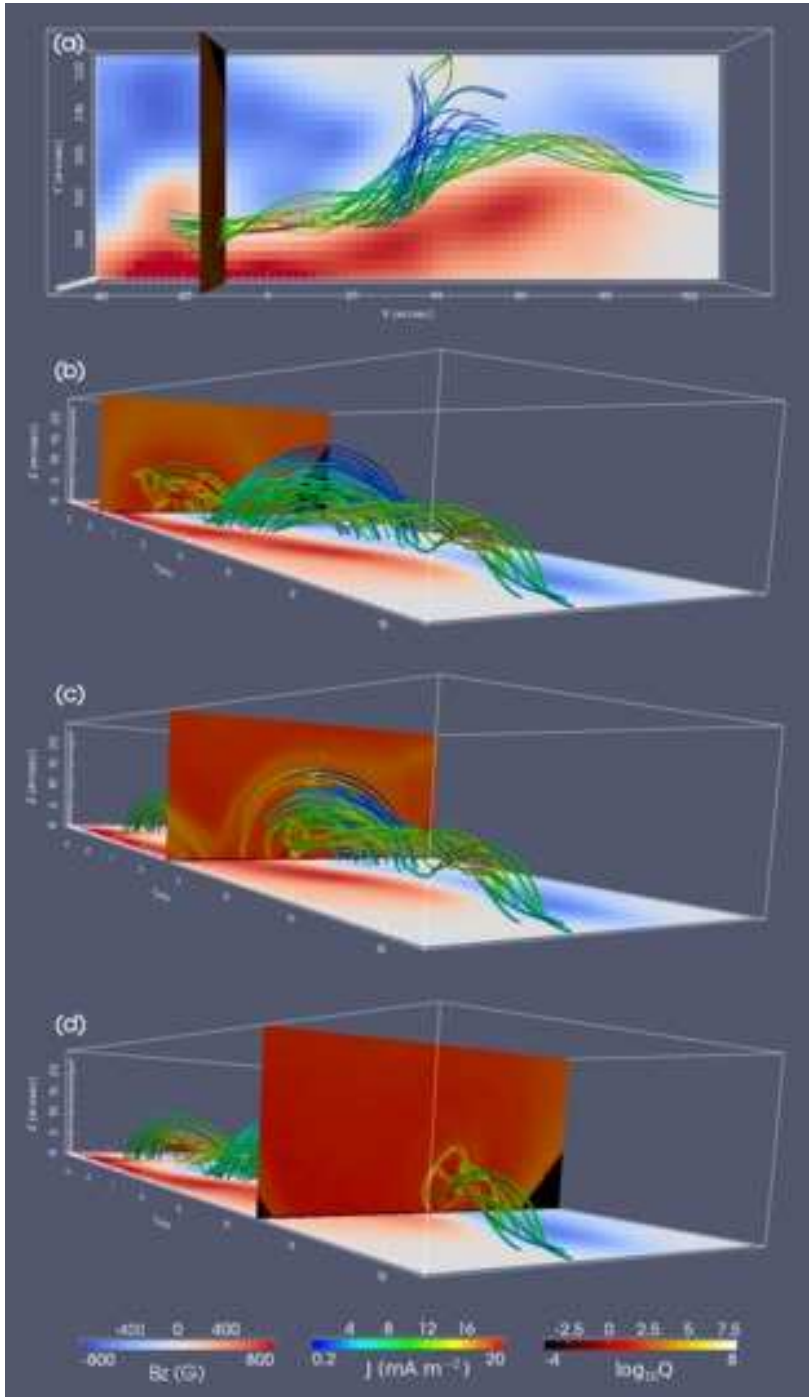}
\caption{QSLs on some sample cuts for the NLFFF at 18:27 UT on 2005 January 15. The three cuts are perpendicular to the $x$-axis and at $x=-10''$, $35''$, and $71''$ for panels (a)/(b), (c), and (d), respectively. The solid lines represent the magnetic field lines, which are rendered with the magnitude of the electric current density ($J$). The image parallel to the $xy$-plane shows $B_z$ at $z=0$, and the images parallel to the $yz$-plane show the distributions of $\log_{10}Q$. The color scales for $B_z$, $J$, and $\log_{10}Q$ are shown by the color bars in the lower part of the figure.
(An animation showing the scanning of the vertical cut is available in the online journal.)} \label{fig:qsl}
\end{figure}

If we plot the QSL cuts in the NLFFF as shown in Figure~\ref{fig:qsl}, we can find clearly what magnetic field structures are associated with the QSLs. An online only movie is attached to Figure~\ref{fig:qsl} to show the scanning of the QSL cut along the $x$-axis. It is found that the self-closed QSLs are located at the border of the magnetic flux rope. In addition to these QSLs, some other QSL sections are associated with highly sheared but not twisted field lines. A twisted field line points to the inverse direction as a potential field would do at the associated magnetic dips. This kind of QSLs can be found, for example, in Figures~\ref{fig:qsl_j2}b and \ref{fig:qsl}c as those QSL sections extend outside the flux rope. There are also some QSL sections inside the flux rope as shown in Figures~\ref{fig:qsl_j2}c and \ref{fig:qsl}d. They are surrounded by the self-closed QSL sections and the electric current density is large there.

\section{Summary and Discussions} \label{sec:disc}

Two major CMEs and the associated flares, M8.6 at 06:38 UT and X2.6 at 23:02 UT, occurred on 2005 January 15 in the fast evolving active region NOAA 10720. In a previous paper, \citet{2010Cheng} found that a magnetic flux rope existed about 5 hours before the X2.6 flare by the NLFFF model. In this paper, we further study the pre-flare brightening by the \textit{SOHO}/EIT and compute the magnetic helicity injection, the vector velocity field, the twist accumulation, and the topology structure of the magnetic flux rope.

First, the helicity flux density is computed via the expression, $G_\theta$, proposed by \citet{2005Pariat} and the velocity field is derived by the DAVE4VM method \citep{2008Schuck} using the time series of vector magnetic fields observed by BBSO/DVMG. Next, we compute the twist of the flux rope using the NLFFF model and the twist density formula proposed by \citet{2006Berger}. We have got the NLFFF models at eight time points during about five hours before the X2.6 flare at 23:02 UT. Thus, the time evolution of the twist is also derived in this period. Finally, we compute the 3D distributions of the squashing degree, $Q$, via the scan of selected 2D cuts in the 3D domain using the method proposed in \citet{2012Pariat}. The QSLs are 3D volumes where the $Q$ values are large.

With the above analysis, we have the following findings. First, there were five C and M class flares that appeared intermittently in the core field along the polarity inversion line in the five hour period before the X2.6 flare. The core field did not erupt to a CME until the one associated with the X2.6 flare.

Secondly, NOAA 10720 was a fast evolving active region, where large horizontal flows and fast magnetic flux emergence existed. These photospheric motions injected negative magnetic helicity into the corona. The helicity injection rate was $(-16.47 \pm 3.52) \times 10^{40}~\mathrm{Mx}^2 ~\mathrm{hr}^{-1}$ from 18:27 UT to 22:20 UT on 2015 January 15. About 1.8\% of the injected magnetic helicity became the internal helicity of the magnetic flux rope, resulting in the twist increasing with a rate of  $-0.18 \pm 0.08$ Turns hr$^{-1}$. The evolutions of the accumulated magnetic helicity and twist had a good correlation with each other. The correlation coefficient was about 0.6. It is expected that most part of the magnetic helicity in a given magnetic configuration is stored in the mutual helicity. From the theoretical analysis of \citet{2006Demoulin2}, the ratio of the self-helicity to the mutual helicity scales to $1/N$, where $N$ is number of the elementary flux tubes in the magnetic field and it is usually large. Besides, we only compute the self-helicity in one major flux rope, but do not count those in other elementary flux tubes. This may explain the large difference in quantity between the injected magnetic helicity and the self-helicity in the major flux rope, though they are highly correlated. As a test, we also estimate the total relative helicity contained in the computation box with the NLFFF models, using the method proposed by \citet{2013Yang1,2013Yang2}. The preliminary results show that it is overall consistent with the injected magnetic helicity from the photosphere. However, since there are still large errors in the computed total relative magnetic helicity, an affirmative conclusion asks for a detailed error analysis and computations in a larger time range of, say, several days.

Thirdly, we find that the flux rope was wrapped by QSLs, which were associated with bald patches and highly sheared and twisted magnetic field lines. The main feature of the QSLs was a hollow shell, but there were also some QSLs in the shell. The magnitude of the electric current density, $J$, did not have a one to one correspondence to the QSLs. For example, QSLs with large $Q$ were not necessarily associated with large $J$, and the distributions of $Q$ and $J$ are different. The largest $Q$ was located on a shell and their distributions are highly intermittent. While the largest $J$ was located at a center and decreased to its periphery. The features of the QSL and electric current density distributions are similar for all the other 7 time points.

\citet{2012Savcheva750,2012Savcheva744} made a detailed comparison between the electric current density and the QSL distributions for a sigmoid observed in 2007 February. The NLFFF extrapolation was made using the flux rope insertion method. Similar results are reached regarding the QSL distributions in \citet{2012Savcheva750,2012Savcheva744} and our studies. Both \citet{2012Savcheva750,2012Savcheva744} and this work find that QSLs wrap the boundary of a flux rope. QSLs and strong currents do not necessarily match each other. There are QSLs without strong electric currents. For example, QSLs could appear in a potential magnetic field with no electric currents, such as the initial configuration of \citet{2005Aulanier}. Strong currents can also exist where no QSLs appear. Here, we have to discriminate the difference between volume currents, which is aligned with the magnetic field, and current sheets, where the electric current is not parallel with the magnetic field. The electric current distribution found here represents the volume currents associated with the twisted magnetic flux rope. While QSls are preferential sites for non-field aligned currents, whose width is very thin. The current sheet is not found in the QSLs derived by the NLFFF model because of two reasons. First, the very small width of a current sheet is out of the resolution ability of the NLFFF model. Second, the NLFFF model intends to smear out the non-field aligned currents (it only allows field aligned currents). The current sheet formation can only be revealed by an MHD process, which has been shown in \citet{2005Aulanier}, \citet{2006Buechner}, and \citet{2009Wilmot-Smith}.

The difference between \citet{2012Savcheva750,2012Savcheva744} and our work lies mainly in the electric current density distributions. While \citet{2012Savcheva750,2012Savcheva744} found that the large electric current is distributed on a hollow shell (see also \citealt{2008Bobra,2011Su}), we find that it is located in a center region and decreases to its periphery. Despite different NLFFF extrapolation methods adopted by \citet{2012Savcheva750,2012Savcheva744} and us, the different distributions of the electric current density may reflect the different nature of the active regions, the former of which is a decaying active region and the latter is a new flux emerging region (private communication with B. Kliem).

\citet{2012Savcheva750,2012Savcheva744} also found a hyperbolic flux tube (HFT), where two QSLs intersect with each other and the $Q$ value is very large. We have made time series of the 3D magnetic field and topology analysis at 8 time points. The last time is about 40 minutes before the flux rope eruption. We do not find HFT along the flux rope in all the time series of the 3D magnetic fields. The difference may either reflect the real structures of the different active regions, or, come from the different extrapolation methods. In \citet{2012Savcheva750,2012Savcheva744}, on the one hand, there was a clear sigmoid in the active region; on the other hand, the flux rope is ejected (not in equilibrium) and relaxes to a solution more or less close to the boundary. Thus, the presence of a flux rope is an initial assumption. Therefore, HFT and bald patches are easy to appear. While with the optimization method, depending on the preprocessing parameters, it tends to smooth the currents and therefore reduce the size and strength of the flux rope.

Based on the above results, we propose the following scenario for the occurrence of the five confined flares and the X2.6 class eruptive flare. The photospheric motions caused by the magnetic flux emergence injected magnetic helicity (and energy, which is not studied here) continuously into the solar corona. Although only a small part ($\sim$ 1.8\%) of the magnetic helicity was built into the magnetic twist of the flux rope, QSLs were created around and inside it. According to previous studies \citep[e.g.,][]{2005Aulanier,2006Buechner,2009Wilmot-Smith}, electric currents are preferentially built up in QSLs. Once the strength of the electric currents were large enough to trigger the resistive instability, magnetic energy in the current layers was released by the magnetic reconnection, which also redistributed the magnetic helicity in the active region. The intermittent confined flares were the manifestation of this reconnection process. We find that the twist of the flux rope always exceeded the critical value of 1.5 and 1.75 turns for the kink instability found by \citet{2003Fan} and \citet{2004Torok}, respectively. However, the flux rope touched the photosphere in bald patches, where the line tied effects prevented the flux rope from an eruption via the kink instability. Consequently, the flux rope did not rise to a height to trigger the torus instability that would lead to a full eruption. For the X2.6 eruptive flare, the mechanism was most probably the same as that proposed in \citet{2012Savcheva744}. An HFT would be formed below the flux rope. Tether-cutting magnetic reconnection \citep{2001Moore} is supposed to occur in the HFT. The flux rope would erupt into the interplanetary space successfully via the loss-of-equilibrium \citep{1991Forbes} or the torus instability \citep{2006Kliem,2010Olmedo}.

\acknowledgments
The authors thank B. Kliem and S. Yang for helpful discussions and thank the referee for constructive suggestions. YG, MDD, and XC were supported by the National Natural Science Foundation of China (NSFC) under the grant numbers 11203014, 10933003, 11373023, 11303016, and the grant from the 973 project 2011CB811402.



\begin{thebibliography}{100}
\expandafter\ifx\csname natexlab\endcsname\relax\def\natexlab#1{#1}\fi

\bibitem[{{Amari} {et~al.}(2003){Amari}, {Luciani}, {Aly}, {Mikic}, \&
  {Linker}}]{2003Amari}
{Amari}, T., {Luciani}, J.~F., {Aly}, J.~J., {Mikic}, Z., \& {Linker}, J. 2003,
  \apj, 595, 1231

\bibitem[{{Amari} {et~al.}(2000){Amari}, {Luciani}, {Mikic}, \&
  {Linker}}]{2000Amari}
{Amari}, T., {Luciani}, J.~F., {Mikic}, Z., \& {Linker}, J. 2000, \apjl, 529,
  L49

\bibitem[{{Antiochos} {et~al.}(1999){Antiochos}, {DeVore}, \&
  {Klimchuk}}]{1999Antiochos}
{Antiochos}, S.~K., {DeVore}, C.~R., \& {Klimchuk}, J.~A. 1999, \apj, 510, 485

\bibitem[{{Aulanier} {et~al.}(2005){Aulanier}, {Pariat}, \&
  {D{\'e}moulin}}]{2005Aulanier}
{Aulanier}, G., {Pariat}, E., \& {D{\'e}moulin}, P. 2005, \aap, 444, 961

\bibitem[{{Aulanier} {et~al.}(2010){Aulanier}, {T{\"o}r{\"o}k}, {D{\'e}moulin},
  \& {DeLuca}}]{2010Aulanier}
{Aulanier}, G., {T{\"o}r{\"o}k}, T., {D{\'e}moulin}, P., \& {DeLuca}, E.~E.
  2010, \apj, 708, 314

\bibitem[{{Berger}(1984)}]{1984Berger2}
{Berger}, M.~A. 1984, \textit{Geophysical and Astrophysical Fluid Dynamics},
  30, 79

\bibitem[{{Berger} \& {Field}(1984)}]{1984Berger}
{Berger}, M.~A. \& {Field}, G.~B. 1984, \textit{Journal of Fluid Mechanics},
  147, 133

\bibitem[{{Berger} \& {Prior}(2006)}]{2006Berger}
{Berger}, M.~A. \& {Prior}, C. 2006, Journal of Physics A Mathematical General,
  39, 8321

\bibitem[{{Bobra} {et~al.}(2008){Bobra}, {van Ballegooijen}, \&
  {DeLuca}}]{2008Bobra}
{Bobra}, M.~G., {van Ballegooijen}, A.~A., \& {DeLuca}, E.~E. 2008, \apj, 672,
  1209

\bibitem[{{Brueckner} {et~al.}(1995){Brueckner}, {Howard}, {Koomen},
  {Korendyke}, {Michels}, {Moses}, {et~al.}}]{1995Brueckner}
{Brueckner}, G.~E., {Howard}, R.~A., {Koomen}, M.~J., {Korendyke}, C.~M.,
  {Michels}, D.~J., {Moses}, J.~D., {et~al.} 1995, \solphys, 162, 357

\bibitem[{{B{\"u}chner}(2006)}]{2006Buechner}
{B{\"u}chner}, J. 2006, \ssr, 122, 149

\bibitem[{{Canou} \& {Amari}(2010)}]{2010Canou}
{Canou}, A. \& {Amari}, T. 2010, \apj, 715, 1566

\bibitem[{{Canou} {et~al.}(2009){Canou}, {Amari}, {Bommier}, {Schmieder},
  {Aulanier}, \& {Li}}]{2009Canou}
{Canou}, A., {Amari}, T., {Bommier}, V., {Schmieder}, B., {Aulanier}, G., \&
  {Li}, H. 2009, \apjl, 693, L27

\bibitem[{{Chae}(2001)}]{2001Chae}
{Chae}, J. 2001, \apjl, 560, L95

\bibitem[{{Chae} {et~al.}(2004){Chae}, {Moon}, \& {Park}}]{2004Chae}
{Chae}, J., {Moon}, Y., \& {Park}, Y. 2004, \solphys, 223, 39

\bibitem[{{Chae} \& {Sakurai}(2008)}]{2008Chae}
{Chae}, J. \& {Sakurai}, T. 2008, \apj, 689, 593

\bibitem[{{Chae} {et~al.}(2001){Chae}, {Wang}, {Qiu}, {Goode}, {Strous}, \&
  {Yun}}]{2001Chae2}
{Chae}, J., {Wang}, H., {Qiu}, J., {Goode}, P.~R., {Strous}, L., \& {Yun},
  H.~S. 2001, \apj, 560, 476

\bibitem[{{Cheng} {et~al.}(2010){Cheng}, {Ding}, {Guo}, {Zhang}, {Jing}, \&
  {Wiegelmann}}]{2010Cheng}
{Cheng}, X., {Ding}, M.~D., {Guo}, Y., {Zhang}, J., {Jing}, J., \&
  {Wiegelmann}, T. 2010, \apjl, 716, L68

\bibitem[{{Cheng} {et~al.}(2013){Cheng}, {Zhang}, {Ding}, {Liu}, \&
  {Poomvises}}]{2013Cheng}
{Cheng}, X., {Zhang}, J., {Ding}, M.~D., {Liu}, Y., \& {Poomvises}, W. 2013,
  \apj, 763, 43

\bibitem[{{Cheng} {et~al.}(2011){Cheng}, {Zhang}, {Liu}, \& {Ding}}]{2011Cheng}
{Cheng}, X., {Zhang}, J., {Liu}, Y., \& {Ding}, M.~D. 2011, \apjl, 732, L25

\bibitem[{{Delaboudini{\`e}re} {et~al.}(1995){Delaboudini{\`e}re}, {Artzner},
  {Brunaud}, {Gabriel}, {Hochedez}, {Millier}, {et~al.}}]{1995Delaboudiniere}
{Delaboudini{\`e}re}, J., {Artzner}, G.~E., {Brunaud}, J., {Gabriel}, A.~H.,
  {Hochedez}, J.~F., {Millier}, F., {et~al.} 1995, \solphys, 162, 291

\bibitem[{{D{\'e}moulin}(2006)}]{2006Demoulin}
{D{\'e}moulin}, P. 2006, Advances in Space Research, 37, 1269

\bibitem[{{D{\'e}moulin}(2007)}]{2007Demoulin}
---. 2007, Advances in Space Research, 39, 1367

\bibitem[{{D{\'e}moulin} \& {Aulanier}(2010)}]{2010Demoulin}
{D{\'e}moulin}, P. \& {Aulanier}, G. 2010, \apj, 718, 1388

\bibitem[{{D{\'e}moulin} \& {Berger}(2003)}]{2003Demoulin}
{D{\'e}moulin}, P. \& {Berger}, M.~A. 2003, \solphys, 215, 203

\bibitem[{{D\'emoulin} {et~al.}(1996){D\'emoulin}, {Henoux}, {Priest}, \&
  {Mandrini}}]{1996Demoulin}
{D\'emoulin}, P., {Henoux}, J.~C., {Priest}, E.~R., \& {Mandrini}, C.~H. 1996,
  \aap, 308, 643

\bibitem[{{D{\'e}moulin} {et~al.}(2006){D{\'e}moulin}, {Pariat}, \&
  {Berger}}]{2006Demoulin2}
{D{\'e}moulin}, P., {Pariat}, E., \& {Berger}, M.~A. 2006, \solphys, 233, 3

\bibitem[{{DeRosa} {et~al.}(2009){DeRosa}, {Schrijver}, {Barnes}, {Leka},
  {Lites}, {Aschwanden}, {et~al.}}]{2009DeRosa}
{DeRosa}, M.~L., {Schrijver}, C.~J., {Barnes}, G., {Leka}, K.~D., {Lites},
  B.~W., {Aschwanden}, M.~J., {et~al.} 2009, \apj, 696, 1780

\bibitem[{{Fan} \& {Gibson}(2003)}]{2003Fan}
{Fan}, Y. \& {Gibson}, S.~E. 2003, \apjl, 589, L105

\bibitem[{{Fan} \& {Gibson}(2004)}]{2004Fan}
---. 2004, \apj, 609, 1123

\bibitem[{{Fan} \& {Gibson}(2007)}]{2007Fan}
---. 2007, \apj, 668, 1232

\bibitem[{{Forbes} \& {Isenberg}(1991)}]{1991Forbes}
{Forbes}, T.~G. \& {Isenberg}, P.~A. 1991, \apj, 373, 294

\bibitem[{{Gary}(2001)}]{2001Gary}
{Gary}, G.~A. 2001, \solphys, 203, 71

\bibitem[{{Georgoulis} \& {LaBonte}(2006)}]{2006Georgoulis}
{Georgoulis}, M.~K. \& {LaBonte}, B.~J. 2006, \apj, 636, 475

\bibitem[{{Green} \& {Kliem}(2009)}]{2009Green}
{Green}, L.~M. \& {Kliem}, B. 2009, \apjl, 700, L83

\bibitem[{{Green} {et~al.}(2011){Green}, {Kliem}, \& {Wallace}}]{2011Green}
{Green}, L.~M., {Kliem}, B., \& {Wallace}, A.~J. 2011, \aap, 526, A2

\bibitem[{{Guo} {et~al.}(2012){Guo}, {Ding}, {Schmieder}, {D{\'e}moulin}, \&
  {Li}}]{2012Guo}
{Guo}, Y., {Ding}, M.~D., {Schmieder}, B., {D{\'e}moulin}, P., \& {Li}, H.
  2012, \apj, 746, 17

\bibitem[{{Guo} {et~al.}(2010{\natexlab{a}}){Guo}, {Ding}, {Schmieder}, {Li},
  {T{\"o}r{\"o}k}, \& {Wiegelmann}}]{2010Guo2}
{Guo}, Y., {Ding}, M.~D., {Schmieder}, B., {Li}, H., {T{\"o}r{\"o}k}, T., \&
  {Wiegelmann}, T. 2010{\natexlab{a}}, \apjl, 725, L38

\bibitem[{{Guo} {et~al.}(2010{\natexlab{b}}){Guo}, {Schmieder}, {D{\'e}moulin},
  {Wiegelmann}, {Aulanier}, {T{\"o}r{\"o}k}, \& {Bommier}}]{2010Guo1}
{Guo}, Y., {Schmieder}, B., {D{\'e}moulin}, P., {Wiegelmann}, T., {Aulanier},
  G., {T{\"o}r{\"o}k}, T., \& {Bommier}, V. 2010{\natexlab{b}}, \apj, 714, 343

\bibitem[{{Huang} {et~al.}(2011){Huang}, {D{\'e}moulin}, {Pick}, {Auch{\`e}re},
  {Yan}, \& {Bouteille}}]{2011Huang}
{Huang}, J., {D{\'e}moulin}, P., {Pick}, M., {Auch{\`e}re}, F., {Yan}, Y.~H.,
  \& {Bouteille}, A. 2011, \apj, 729, 107

\bibitem[{{Ji} {et~al.}(2003){Ji}, {Wang}, {Schmahl}, {Moon}, \&
  {Jiang}}]{2003Ji}
{Ji}, H., {Wang}, H., {Schmahl}, E.~J., {Moon}, Y., \& {Jiang}, Y. 2003, \apjl,
  595, L135

\bibitem[{{Jing} {et~al.}(2009){Jing}, {Chen}, {Wiegelmann}, {Xu}, {Park}, \&
  {Wang}}]{2009Jing}
{Jing}, J., {Chen}, P.~F., {Wiegelmann}, T., {Xu}, Y., {Park}, S., \& {Wang},
  H. 2009, \apj, 696, 84

\bibitem[{{Jing} {et~al.}(2012){Jing}, {Park}, {Liu}, {Lee}, {Wiegelmann},
  {Xu}, {Deng}, \& {Wang}}]{2012Jing}
{Jing}, J., {Park}, S.-H., {Liu}, C., {Lee}, J., {Wiegelmann}, T., {Xu}, Y.,
  {Deng}, N., \& {Wang}, H. 2012, \apjl, 752, L9

\bibitem[{{Judge}(1998)}]{1998Judge}
{Judge}, P.~G. 1998, \apj, 500, 1009

\bibitem[{{Kliem} \& {T{\"o}r{\"o}k}(2006)}]{2006Kliem}
{Kliem}, B. \& {T{\"o}r{\"o}k}, T. 2006, Physical Review Letters, 96, 255002

\bibitem[{{Kuckein} {et~al.}(2009){Kuckein}, {Centeno}, {Mart{\'{\i}}nez
  Pillet}, {Casini}, {Manso Sainz}, \& {Shimizu}}]{2009Kuckein}
{Kuckein}, C., {Centeno}, R., {Mart{\'{\i}}nez Pillet}, V., {Casini}, R.,
  {Manso Sainz}, R., \& {Shimizu}, T. 2009, \aap, 501, 1113

\bibitem[{{Kuckein} {et~al.}(2012){Kuckein}, {Mart{\'{\i}}nez Pillet}, \&
  {Centeno}}]{2012Kuckein}
{Kuckein}, C., {Mart{\'{\i}}nez Pillet}, V., \& {Centeno}, R. 2012, \aap, 539,
  A131

\bibitem[{{Kusano} {et~al.}(2002){Kusano}, {Maeshiro}, {Yokoyama}, \&
  {Sakurai}}]{2002Kusano}
{Kusano}, K., {Maeshiro}, T., {Yokoyama}, T., \& {Sakurai}, T. 2002, \apj, 577,
  501

\bibitem[{{Kusano} {et~al.}(2004){Kusano}, {Maeshiro}, {Yokoyama}, \&
  {Sakurai}}]{2004Kusano}
{Kusano}, K., {Maeshiro}, T., {Yokoyama}, T., \& {Sakurai}, T. 2004, in ASP
  Conf. Ser., Vol. 325, The Solar-B Mission and the Forefront of Solar Physics,
  ed. {T.~Sakurai \& T.~Sekii}, 175

\bibitem[{{Leka} {et~al.}(2009){Leka}, {Barnes}, {Crouch}, {Metcalf}, {Gary},
  {Jing}, \& {Liu}}]{2009Leka}
{Leka}, K.~D., {Barnes}, G., {Crouch}, A.~D., {Metcalf}, T.~R., {Gary}, G.~A.,
  {Jing}, J., \& {Liu}, Y. 2009, \solphys, 260, 83

\bibitem[{{Li} \& {Zhang}(2013{\natexlab{a}})}]{2013Li1}
{Li}, L.~P. \& {Zhang}, J. 2013{\natexlab{a}}, \aap, 552, L11

\bibitem[{{Li} \& {Zhang}(2013{\natexlab{b}})}]{2013Li2}
{Li}, T. \& {Zhang}, J. 2013{\natexlab{b}}, ArXiv e-prints

\bibitem[{{Lim} {et~al.}(2007){Lim}, {Jeong}, {Chae}, \& {Moon}}]{2007Lim}
{Lim}, E.-K., {Jeong}, H., {Chae}, J., \& {Moon}, Y.-J. 2007, \apj, 656, 1167

\bibitem[{{Lin} {et~al.}(2000){Lin}, {Penn}, \& {Tomczyk}}]{2000Lin}
{Lin}, H., {Penn}, M.~J., \& {Tomczyk}, S. 2000, \apjl, 541, L83

\bibitem[{{Liu} {et~al.}(2010){Liu}, {Liu}, {Wang}, {Deng}, \&
  {Wang}}]{2010Liu}
{Liu}, R., {Liu}, C., {Wang}, S., {Deng}, N., \& {Wang}, H. 2010, \apjl, 725,
  L84

\bibitem[{{Longcope}(2004)}]{2004Longcope}
{Longcope}, D.~W. 2004, \apj, 612, 1181

\bibitem[{{Metcalf}(1994)}]{1994Metcalf}
{Metcalf}, T.~R. 1994, \solphys, 155, 235

\bibitem[{{Metcalf} {et~al.}(2008){Metcalf}, {De Rosa}, {Schrijver}, {Barnes},
  {van Ballegooijen}, {Wiegelmann}, {Wheatland}, {Valori}, \&
  {McTtiernan}}]{2008Metcalf}
{Metcalf}, T.~R., {De Rosa}, M.~L., {Schrijver}, C.~J., {Barnes}, G., {van
  Ballegooijen}, A.~A., {Wiegelmann}, T., {Wheatland}, M.~S., {Valori}, G., \&
  {McTtiernan}, J.~M. 2008, \solphys, 247, 269

\bibitem[{{Metcalf} {et~al.}(2006){Metcalf}, {Leka}, {Barnes}, {Lites},
  {Georgoulis}, {Pevtsov}, {et~al.}}]{2006Metcalf}
{Metcalf}, T.~R., {Leka}, K.~D., {Barnes}, G., {Lites}, B.~W., {Georgoulis},
  M.~K., {Pevtsov}, A.~A., {et~al.} 2006, \solphys, 237, 267

\bibitem[{{Moon} {et~al.}(2002{\natexlab{a}}){Moon}, {Chae}, {Choe}, {Wang},
  {Park}, {Yun}, {Yurchyshyn}, \& {Goode}}]{2002Moon1}
{Moon}, Y., {Chae}, J., {Choe}, G.~S., {Wang}, H., {Park}, Y.~D., {Yun}, H.~S.,
  {Yurchyshyn}, V., \& {Goode}, P.~R. 2002{\natexlab{a}}, \apj, 574, 1066

\bibitem[{{Moon} {et~al.}(2002{\natexlab{b}}){Moon}, {Chae}, {Wang}, {Choe}, \&
  {Park}}]{2002Moon2}
{Moon}, Y., {Chae}, J., {Wang}, H., {Choe}, G.~S., \& {Park}, Y.~D.
  2002{\natexlab{b}}, \apj, 580, 528

\bibitem[{{Moore} \& {Roumeliotis}(1992)}]{1992Moore}
{Moore}, R.~L. \& {Roumeliotis}, G. 1992, in Lecture Notes in Physics, Berlin
  Springer Verlag, Vol. 399, IAU Colloq. 133: Eruptive Solar Flares, ed.
  Z.~{Svestka}, B.~V. {Jackson}, \& M.~E. {Machado}, 69

\bibitem[{{Moore} {et~al.}(2001){Moore}, {Sterling}, {Hudson}, \&
  {Lemen}}]{2001Moore}
{Moore}, R.~L., {Sterling}, A.~C., {Hudson}, H.~S., \& {Lemen}, J.~R. 2001,
  \apj, 552, 833

\bibitem[{{Nindos} \& {Zhang}(2002)}]{2002Nindos}
{Nindos}, A. \& {Zhang}, H. 2002, \apjl, 573, L133

\bibitem[{{Nindos} {et~al.}(2003){Nindos}, {Zhang}, \& {Zhang}}]{2003Nindos}
{Nindos}, A., {Zhang}, J., \& {Zhang}, H. 2003, \apj, 594, 1033

\bibitem[{{Olmedo} \& {Zhang}(2010)}]{2010Olmedo}
{Olmedo}, O. \& {Zhang}, J. 2010, \apj, 718, 433

\bibitem[{{Pariat} \& {D{\'e}moulin}(2012)}]{2012Pariat}
{Pariat}, E. \& {D{\'e}moulin}, P. 2012, \aap, 541, A78

\bibitem[{{Pariat} {et~al.}(2005){Pariat}, {D{\'e}moulin}, \&
  {Berger}}]{2005Pariat}
{Pariat}, E., {D{\'e}moulin}, P., \& {Berger}, M.~A. 2005, \aap, 439, 1191

\bibitem[{{Park} {et~al.}(2010){Park}, {Chae}, {Jing}, {Tan}, \&
  {Wang}}]{2010Park}
{Park}, S.-H., {Chae}, J., {Jing}, J., {Tan}, C., \& {Wang}, H. 2010, \apj,
  720, 1102

\bibitem[{{Patsourakos} {et~al.}(2013){Patsourakos}, {Vourlidas}, \&
  {Stenborg}}]{2013Patsourakos}
{Patsourakos}, S., {Vourlidas}, A., \& {Stenborg}, G. 2013, \apj, 764, 125

\bibitem[{{Priest} \& {Forbes}(2000)}]{2000Priest}
{Priest}, E. \& {Forbes}, T., eds. 2000, {Magnetic reconnection : MHD theory
  and applications}

\bibitem[{{Priest} \& {D{\'e}moulin}(1995)}]{1995Priest}
{Priest}, E.~R. \& {D{\'e}moulin}, P. 1995, \jgr, 100, 23443

\bibitem[{{R{\'e}gnier} {et~al.}(2002){R{\'e}gnier}, {Amari}, \&
  {Kersal{\'e}}}]{2002Regnier}
{R{\'e}gnier}, S., {Amari}, T., \& {Kersal{\'e}}, E. 2002, \aap, 392, 1119

\bibitem[{{Roussev} {et~al.}(2003){Roussev}, {Forbes}, {Gombosi}, {Sokolov},
  {DeZeeuw}, \& {Birn}}]{2003roussev}
{Roussev}, I.~I., {Forbes}, T.~G., {Gombosi}, T.~I., {Sokolov}, I.~V.,
  {DeZeeuw}, D.~L., \& {Birn}, J. 2003, \apjl, 588, L45

\bibitem[{{Rudenko} \& {Myshyakov}(2011)}]{2011Rudenko}
{Rudenko}, G.~V. \& {Myshyakov}, I.~I. 2011, \solphys, 270, 165

\bibitem[{{Savcheva} {et~al.}(2012{\natexlab{a}}){Savcheva}, {Pariat}, {van
  Ballegooijen}, {Aulanier}, \& {DeLuca}}]{2012Savcheva750}
{Savcheva}, A., {Pariat}, E., {van Ballegooijen}, A., {Aulanier}, G., \&
  {DeLuca}, E. 2012{\natexlab{a}}, \apj, 750, 15

\bibitem[{{Savcheva} \& {van Ballegooijen}(2009)}]{2009Savcheva}
{Savcheva}, A. \& {van Ballegooijen}, A. 2009, \apj, 703, 1766

\bibitem[{{Savcheva} {et~al.}(2012{\natexlab{b}}){Savcheva}, {van
  Ballegooijen}, \& {DeLuca}}]{2012Savcheva744}
{Savcheva}, A., {van Ballegooijen}, A., \& {DeLuca}, E. 2012{\natexlab{b}},
  \apj, 744, 78

\bibitem[{{Schrijver} {et~al.}(2006){Schrijver}, {De Rosa}, {Metcalf}, {Liu},
  {McTiernan}, {R{\'e}gnier}, {Valori}, {Wheatland}, \&
  {Wiegelmann}}]{2006Schrijver}
{Schrijver}, C.~J., {De Rosa}, M.~L., {Metcalf}, T.~R., {Liu}, Y., {McTiernan},
  J., {R{\'e}gnier}, S., {Valori}, G., {Wheatland}, M.~S., \& {Wiegelmann}, T.
  2006, \solphys, 235, 161

\bibitem[{{Schuck}(2005)}]{2005Schuck}
{Schuck}, P.~W. 2005, \apjl, 632, L53

\bibitem[{{Schuck}(2006)}]{2006Schuck}
---. 2006, \apj, 646, 1358

\bibitem[{{Schuck}(2008)}]{2008Schuck}
---. 2008, \apj, 683, 1134

\bibitem[{{Su} {et~al.}(2011){Su}, {Surges}, {van Ballegooijen}, {DeLuca}, \&
  {Golub}}]{2011Su}
{Su}, Y., {Surges}, V., {van Ballegooijen}, A., {DeLuca}, E., \& {Golub}, L.
  2011, \apj, 734, 53

\bibitem[{{Thalmann} {et~al.}(2011){Thalmann}, {Inhester}, \&
  {Wiegelmann}}]{2011Thalmann}
{Thalmann}, J.~K., {Inhester}, B., \& {Wiegelmann}, T. 2011, \solphys, 272, 243

\bibitem[{{Titov} {et~al.}(2002){Titov}, {Hornig}, \&
  {D{\'e}moulin}}]{2002Titov}
{Titov}, V.~S., {Hornig}, G., \& {D{\'e}moulin}, P. 2002, Journal of
  Geophysical Research (Space Physics), 107, 1164

\bibitem[{{T{\"o}r{\"o}k} \& {Kliem}(2005)}]{2005Torok}
{T{\"o}r{\"o}k}, T. \& {Kliem}, B. 2005, \apjl, 630, L97

\bibitem[{{T{\"o}r{\"o}k} {et~al.}(2004){T{\"o}r{\"o}k}, {Kliem}, \&
  {Titov}}]{2004Torok}
{T{\"o}r{\"o}k}, T., {Kliem}, B., \& {Titov}, V.~S. 2004, \aap, 413, L27

\bibitem[{{Valori} {et~al.}(2012){Valori}, {D{\'e}moulin}, \&
  {Pariat}}]{2012Valori}
{Valori}, G., {D{\'e}moulin}, P., \& {Pariat}, E. 2012, \solphys, 278, 347

\bibitem[{{Wang} {et~al.}(2009){Wang}, {Muglach}, \& {Kliem}}]{2009Wang}
{Wang}, Y., {Muglach}, K., \& {Kliem}, B. 2009, \apj, 699, 133

\bibitem[{{Welsch} {et~al.}(2007){Welsch}, {Abbett}, {De Rosa}, {Fisher},
  {Georgoulis}, {Kusano}, {Longcope}, {Ravindra}, \& {Schuck}}]{2007Welsch}
{Welsch}, B.~T., {Abbett}, W.~P., {De Rosa}, M.~L., {Fisher}, G.~H.,
  {Georgoulis}, M.~K., {Kusano}, K., {Longcope}, D.~W., {Ravindra}, B., \&
  {Schuck}, P.~W. 2007, \apj, 670, 1434

\bibitem[{{Welsch} {et~al.}(2004){Welsch}, {Fisher}, {Abbett}, \&
  {Regnier}}]{2004Welsch}
{Welsch}, B.~T., {Fisher}, G.~H., {Abbett}, W.~P., \& {Regnier}, S. 2004, \apj,
  610, 1148

\bibitem[{{Wheatland} {et~al.}(2000){Wheatland}, {Sturrock}, \&
  {Roumeliotis}}]{2000Wheatland}
{Wheatland}, M.~S., {Sturrock}, P.~A., \& {Roumeliotis}, G. 2000, \apj, 540,
  1150

\bibitem[{{Wiegelmann}(2004)}]{2004Wiegelmann}
{Wiegelmann}, T. 2004, \solphys, 219, 87

\bibitem[{{Wiegelmann} {et~al.}(2006){Wiegelmann}, {Inhester}, \&
  {Sakurai}}]{2006Wiegelmann}
{Wiegelmann}, T., {Inhester}, B., \& {Sakurai}, T. 2006, \solphys, 233, 215

\bibitem[{{Wiegelmann} \& {Neukirch}(2002)}]{2002Wiegelmann}
{Wiegelmann}, T. \& {Neukirch}, T. 2002, \solphys, 208, 233

\bibitem[{{Wiegelmann} \& {Sakurai}(2012)}]{2012Wiegelmann}
{Wiegelmann}, T. \& {Sakurai}, T. 2012, Living Reviews in Solar Physics, 9, 5

\bibitem[{{Wilmot-Smith} {et~al.}(2009){Wilmot-Smith}, {Hornig}, \&
  {Pontin}}]{2009Wilmot-Smith}
{Wilmot-Smith}, A.~L., {Hornig}, G., \& {Pontin}, D.~I. 2009, \apj, 704, 1288

\bibitem[{{Yang} {et~al.}(2013{\natexlab{a}}){Yang}, {B{\"u}chner}, {Santos},
  \& {Zhang}}]{2013Yang1}
{Yang}, S., {B{\"u}chner}, J., {Santos}, J.~C., \& {Zhang}, H.
  2013{\natexlab{a}}, \solphys, 283, 369

\bibitem[{{Yang} {et~al.}(2013{\natexlab{b}}){Yang}, {B{\"u}chner}, {Santos},
  \& {Zhang}}]{2013Yang2}
---. 2013{\natexlab{b}}, ArXiv e-prints

\bibitem[{{Zhang} {et~al.}(2012){Zhang}, {Cheng}, \& {Ding}}]{2012Zhang}
{Zhang}, J., {Cheng}, X., \& {Ding}, M.-D. 2012, Nature Communications, 3

\end{thebibliography}

\clearpage

\end{document}